\documentclass[twocolumn,superscriptaddress,pra,aps,preprintnumbers,nofootinbib,longbibliography]{revtex4-2}
\usepackage{amsmath,amssymb,bm,graphicx,color,gensymb,bbold,appendix}
\usepackage[colorlinks=true, citecolor={blue!80!black}, urlcolor={blue!50!black}, linkcolor = {blue!80!black}]{hyperref}
\usepackage{comment}

\usepackage{xcolor}

\usepackage{pdfpages} 
\usepackage{pgffor} 
\makeatletter
\AtBeginDocument{\let\LS@rot\@undefined}
\makeatother
\pdfximage{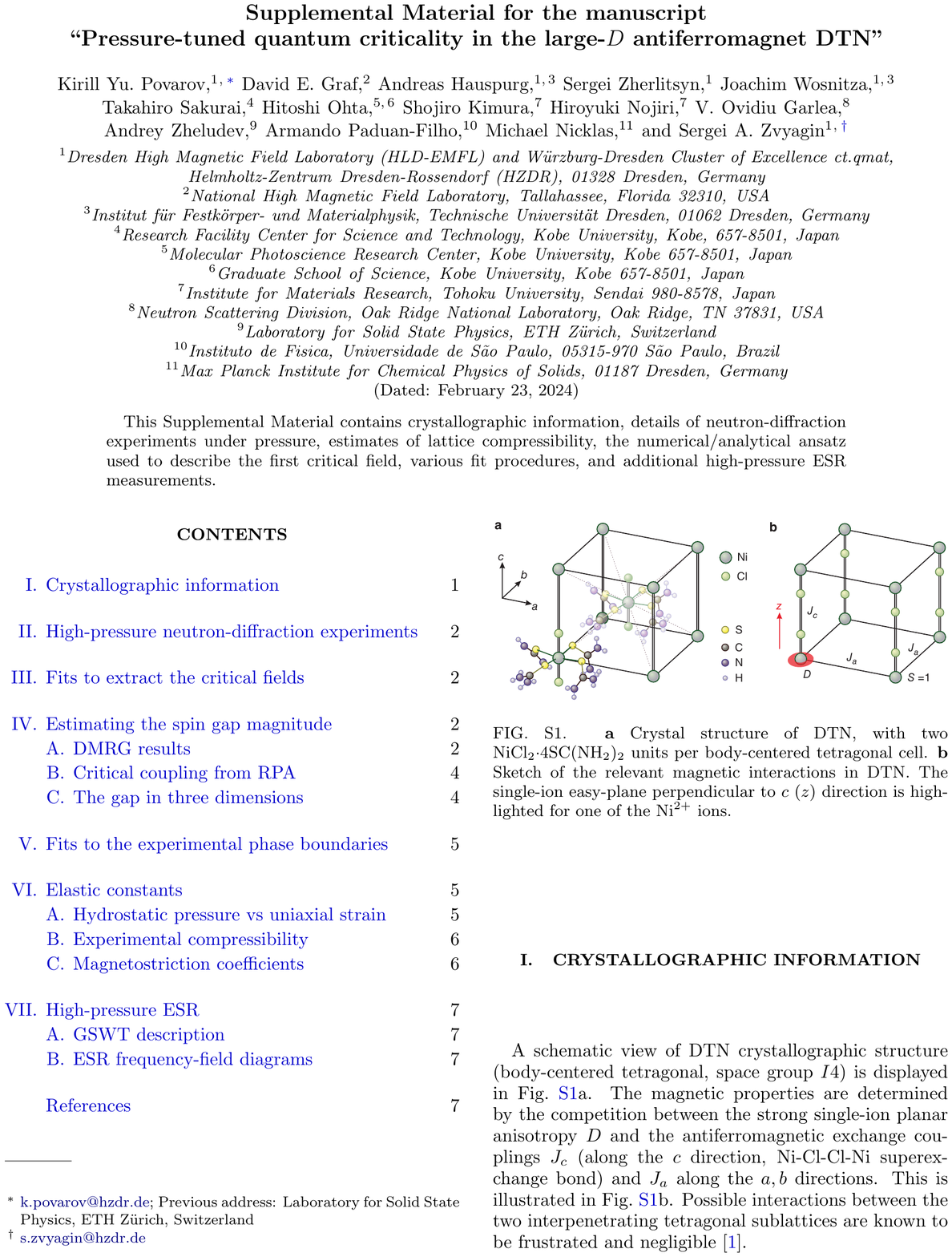}
\def\numbersupplementpages{\the\pdflastximagepages}

\newcommand{\CsFeCl}{CsFeCl$_3$}
\newcommand{\XFeCl}{Cs$_{1-x}$Rb$_{x}$FeCl$_3$}

\newcommand{\hamilt}{\hat{\mathcal{H}}}
\newcommand{\spin}{\hat{S}}
\newcommand{\spop}{\hat{\mathbf{S}}}

\newcommand{\ket}[1]{\left| #1 \right\rangle}

\newcommand{\mub}{\mu_\mathrm{B}}
\newcommand{\kb}{k_\mathrm{B}}

\begin{document}
\title{Pressure-tuned quantum criticality in the large-$D$ antiferromagnet DTN }

\author{Kirill~Yu.~Povarov} 
\email{k.povarov@hzdr.de}
\altaffiliation[Previous address: ]{Laboratory for Solid State Physics, ETH Z\"{u}rich, Switzerland}
\affiliation{Dresden High Magnetic Field Laboratory (HLD-EMFL) and W\"urzburg-Dresden Cluster of Excellence ct.qmat, Helmholtz-Zentrum Dresden-Rossendorf (HZDR), 01328 Dresden, Germany}

\author{David~E.~Graf} 
\affiliation{National High Magnetic Field Laboratory, Tallahassee, Florida 32310, USA}

\author{Andreas~Hauspurg} 
\affiliation{Dresden High Magnetic Field Laboratory (HLD-EMFL) and W\"urzburg-Dresden Cluster of Excellence ct.qmat, Helmholtz-Zentrum Dresden-Rossendorf (HZDR), 01328 Dresden, Germany}
\affiliation{Institut f\"ur Festk\"orper- und Materialphysik, Technische Universit\"at Dresden, 01062 Dresden, Germany}

\author{Sergei~Zherlitsyn} 
\affiliation{Dresden High Magnetic Field Laboratory (HLD-EMFL) and W\"urzburg-Dresden Cluster of Excellence ct.qmat, Helmholtz-Zentrum Dresden-Rossendorf (HZDR), 01328 Dresden, Germany}

\author{Joachim~Wosnitza} 
\affiliation{Dresden High Magnetic Field Laboratory (HLD-EMFL) and W\"urzburg-Dresden Cluster of Excellence ct.qmat, Helmholtz-Zentrum Dresden-Rossendorf (HZDR), 01328 Dresden, Germany}
\affiliation{Institut f\"ur Festk\"orper- und Materialphysik, Technische Universit\"at Dresden, 01062 Dresden, Germany}

\author{Takahiro~Sakurai} 
\affiliation{Research Facility Center for Science and Technology, Kobe University, Kobe, 657-8501, Japan}

\author{Hitoshi~Ohta} 
\affiliation{Molecular Photoscience Research Center, Kobe University, Kobe 657-8501, Japan}
\affiliation{Graduate School of Science, Kobe University, Kobe 657-8501, Japan}

\author{Shojiro~Kimura} 
\affiliation{Institute for Materials Research, Tohoku University, Sendai 980-8578, Japan}

\author{Hiroyuki~Nojiri} 
\affiliation{Institute for Materials Research, Tohoku University, Sendai 980-8578, Japan}

\author{V. Ovidiu Garlea} 
\affiliation{Neutron Scattering Division, Oak Ridge National Laboratory, Oak Ridge, TN 37831, USA}

\author{Andrey~Zheludev} 
\affiliation{Laboratory for Solid State Physics, ETH Z\"{u}rich, Switzerland}

\author{Armando~Paduan-Filho} 
\affiliation{Instituto de Fisica, Universidade de S\~{a}o Paulo,
05315-970 S\~{a}o Paulo, Brazil}

\author{Michael~Nicklas} 
\affiliation{Max Planck Institute for Chemical Physics of Solids, 01187 Dresden, Germany}

\author{Sergei~A.~Zvyagin} 
\email{s.zvyagin@hzdr.de}
\affiliation{Dresden High Magnetic Field Laboratory (HLD-EMFL) and W\"urzburg-Dresden Cluster of Excellence ct.qmat, Helmholtz-Zentrum Dresden-Rossendorf (HZDR), 01328 Dresden, Germany}

\begin{abstract}
Strongly correlated spin systems  can be driven to  quantum critical points   via various routes. In particular,  gapped quantum antiferromagnets can undergo phase transitions into a magnetically ordered state with  applied pressure  or magnetic field, acting as tuning parameters. These transitions are characterized  by $z=1$ or $z=2$ dynamical critical exponents, determined  by the linear and  quadratic  low-energy dispersion of spin excitations, respectively.    Employing  high-frequency susceptibility and ultrasound techniques, we demonstrate that the tetragonal easy-plane quantum antiferromagnet NiCl$_{2}\cdot$4SC(NH$_2$)$_2$ (aka DTN) undergoes a spin-gap closure transition at  about $4.2$ kbar, resulting in a pressure-induced magnetic ordering. The studies are complemented by high-pressure-electron spin-resonance measurements confirming the proposed scenario. Powder neutron diffraction measurements revealed  that no lattice distortion occurs at this pressure and  the high spin symmetry is preserved, establishing  DTN as a perfect platform  to investigate $z=1$ quantum critical phenomena. The experimental observations are  supported by DMRG calculations,
allowing us to  quantitatively describe the pressure-driven evolution of critical fields and spin-Hamiltonian parameters in DTN.

\end{abstract}
\date{\today}
\maketitle

\begin{figure}
 \includegraphics[width=0.5\textwidth]{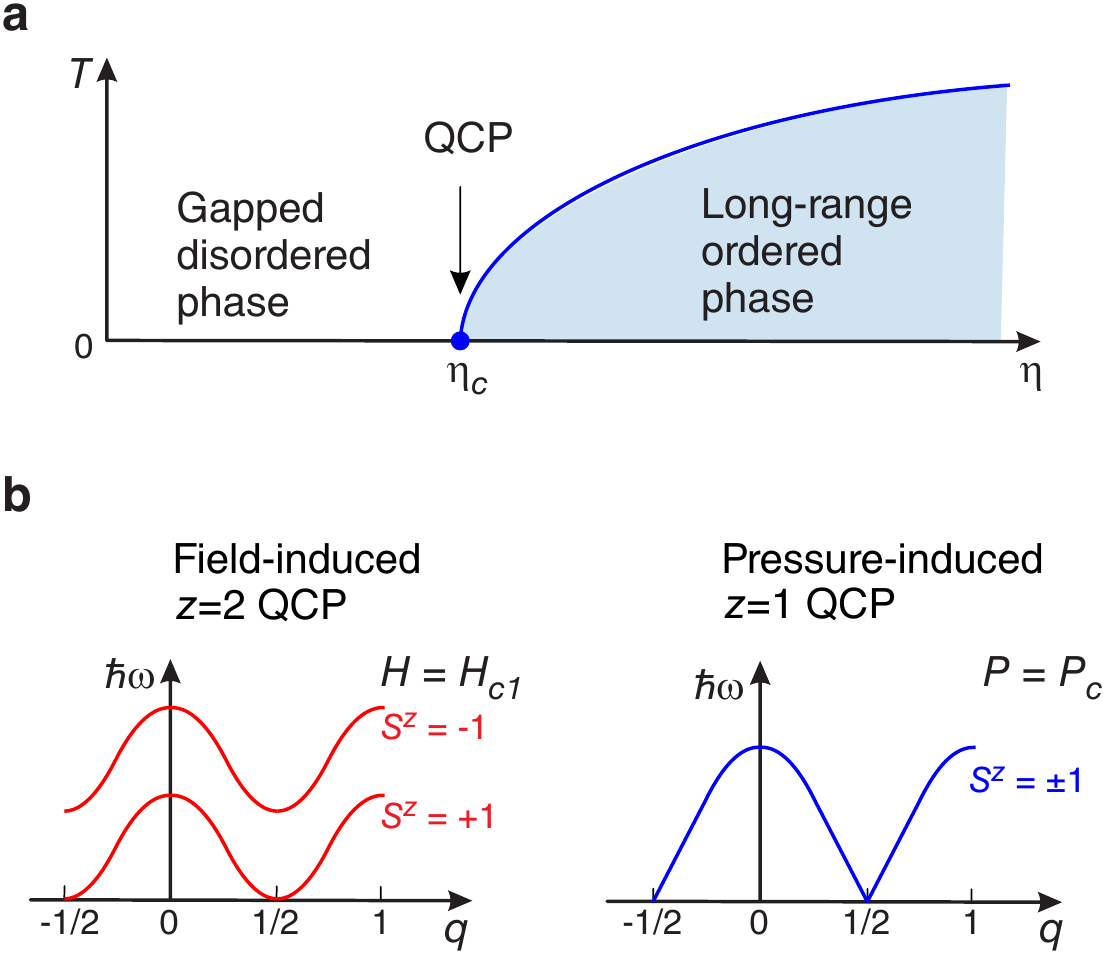}
 \caption{Quantum phase transitions in a  spin-1 AF with large planar anisotropy. \textbf{a} The generic phase diagram ($\eta$ is the tuning parameter; QCP is the quantum critical point).  The critical value of the  tuning parameter is labeled $\eta_c$. \textbf{b} Schematic view of magnetic excitation spectra   for the field- and pressure-induced quantum phase transitions ($q$ is the momentum;  $\hbar\omega$ is the excitation energy; $z$ is the dynamic critical exponent, see Eq.~\ref{EQ:CritZ}).}\label{FIG:intro}
\end{figure}

Magnetic insulators with their short-range interactions  and well-controlled effective Hamiltonians offer an ideal playground for studying quantum critical phenomena induced by various stimuli~\cite{Sachdev_2011_QPTBook,Vojta_RevProgPhys_2003_QCPreview,Sachdev_Science_2000_QCPs}. Nowadays, great attention is attracted by antiferromagnetic (AF) ordering, which  can be  induced, e.g., by magnetic field in gapped quantum  magnets~\cite{Giamarchi_NatPhys_2008_BECreview,Zapf_RMP_2014_BECreview}. On the other hand, especially tantalizing is the possibility of magnetic ordering, accompanied by spontaneous
spin-symmetry break-down   without the aid of magnetic field, while keeping the spectrum degenerate in absence of Zeeman splitting~\cite{Matsumoto_JPSJ_2007_DTNHiggs}. Such transitions are known to appear  in certain quantum antiferromagnets  under applied pressure. As Fig.~\ref{FIG:intro}a illustrates, both the magnetic field and pressure can be regarded as a tuning parameter. Contrary to the field-induced transitions, the pressure-induced case  belongs to a another class of universality ~\cite{MatsumotoNormand_PRB_2004_TlCuClz1z2,ZhangWierschem_PRB_2013_DTNdispersion,ScammellSushkov_PRB_2017_Multiuniversal}. The key distinction is the dynamic critical exponent $z$ that links the characteristic energy of spin excitations  to the momentum relative to the critical one:
\begin{equation}\label{EQ:CritZ}
 \omega\propto |q-q_{c}|^z.
\end{equation}

Field-induced phase transitions are characterized by $z=2$ and a quadratic low-energy excitation spectrum, while the pressure-induced case has $z=1$ and a linear spectrum~\cite{Matsumoto_JPSJ_2007_DTNHiggs,ZhangWierschem_PRB_2013_DTNdispersion}. The realizations of these two scenarios are illustrated in Fig.~\ref{FIG:intro}b. The  $z=1$ regime  is very remarkable. It should have mean-field critical exponents~\cite{ZhangWierschem_PRB_2013_DTNdispersion} and  the universal scaling of dynamic fluctuations at the same time~\cite{Zheludev_JETP_2020_1Dreview}. This quantum critical point (QCP) is also a natural habitat for a well-defined order-parameter amplitude mode~\cite{Ruegg_PRL_2008_PindTlCuCl,Merchant_NatPhys_2014_PindTlCuCl}.

There is a number of gapped quantum antiferromagnets, demonstrating field-induced AF ordering with $z=2$~\cite{Zapf_RMP_2014_BECreview}.
If the spin Hamiltonian of a system
has axial symmetry with respect to the applied field, the field-induced AF
ordering can be formally described  as the Bose-Einstein condensation of magnons by mapping the spin-1 system into a gas of semi-hard-core bosons~\cite{Giamarchi_NatPhys_2008_BECreview, Mills_PRL_2007_BECcomment}. In order to perform accurate comparison with the theory,  high-symmetry spin systems are highly demanded. The  compound NiCl$_{2}\cdot$4SC(NH$_2$)$_2$ (dichloro-tetrakis thiourea-nickel(II), known as DTN) is one of them, having tetragonal crystal structure (space group $I$4) and easily accessible critical fields.

A good realization of the true $z=1$ QCP is actually hard to find. Most of the materials known to show such a transition (TlCuCl$_3$~\cite{Ruegg_PRL_2008_PindTlCuCl,Merchant_NatPhys_2014_PindTlCuCl}, KCuCl$_3$~\cite{GotoFujisawa_JPSJ_2006_KCuCl3press,MatsumotoSakurai_ApplMagRes_2021_KCuCl3Higgs}, (C$_4$H$_{12}$N$_2$)Cu$_2$Cl$_6$~\cite{Thede_PRL_2014_uSRPHCC}, and (C$_9$H$_{18}$N$_2$)CuBr$_4$~\cite{HongYing_NatComm_2022_DLCBpressure}) suffer from unwanted anisotropies. For  TlCuCl$_3$ and (C$_9$H$_{18}$N$_2$)CuBr$_4$ biaxial anisotropy was experimentally detected~\cite{Glazkov_PRB_2004_TlCuCl3gap,GlazkovYankova_PRB_2012_PHCCesr}. In KCuCl$_3$ the presence of biaxial anisotropy follows from symmetry considerations, and is indirectly evidenced by a peculiar orientation of sublattice magnetization in the ordered phase~\cite{GotoFujisawa_JPSJ_2006_KCuCl3press}. Such anisotropies eventually lead to the criticality of the Ising universality class~\cite{DellAmore_PRB_2009_BECbreakdown}, different from the target QCP. This does not seem to be the case for \CsFeCl~\cite{KuritaTanaka_PRB_2016_CsFeCl3pressure,HayashidaMatsumoto_SciAdv_2019_QCPmagnons}, but there one has to deal with geometric exchange frustration.

In the present study, utilizing high-pressure tunneling-diode oscillator (TDO)  susceptibility,   ultrasound-propagation measurements and high-field electron spin resonance (ESR) techniques, we demonstrate a pressure-induced phase transition in DTN, that we ascribe to the long-sought $z=1$ criticality. This transition resides at an easily accessible pressure of about $4.2$~kbar. Neutron-diffraction measurements confirm the absence of a structural transition and reveal an undistorted tetragonal symmetry near this QCP. At higher pressure, we actually find an irreversible distortion of the lattice occurring. We describe the experimentally measured phase boundaries employing a quasi-1D numerical approximation, circumventing a renormalization of the spin-Hamiltonian parameters  by quantum fluctuations.

\begin{figure}
 \includegraphics[width=0.48\textwidth]{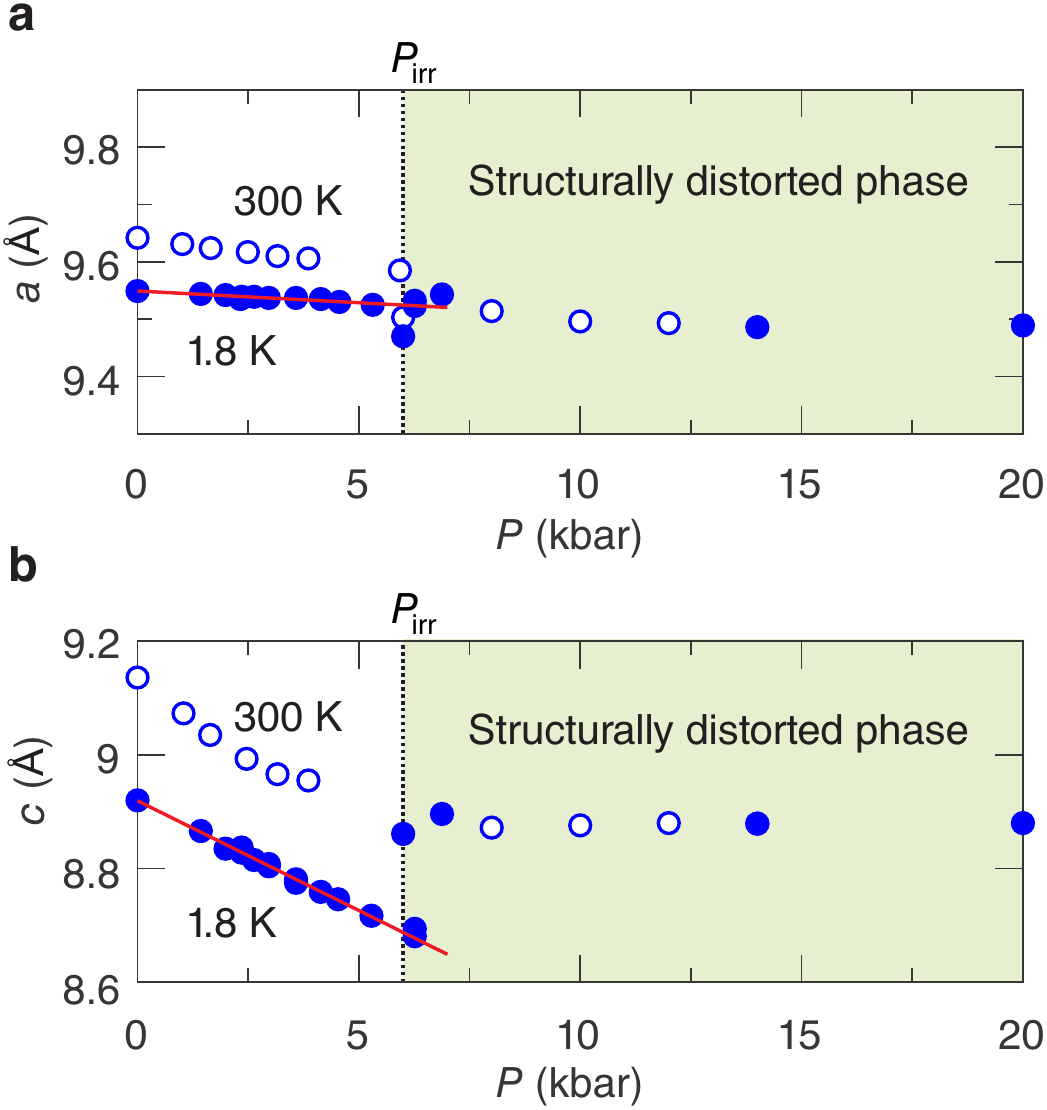}
 \caption{Pressure dependence of  lattice parameters of DTN, as measured by  neutron diffraction. \textbf{a} Pressure dependence of the  lattice parameter $a$.  \textbf{b}  Pressure dependence of the lattice parameter $c$. Open and closed  symbols correspond to 300  and  $1.8$~K data, respectively.   Dotted lines denote the irreversible  structural phase transition at $P_\mathrm{irr} \sim 6$~kbar. Red lines are fit results (see text for details).}\label{FIG:lattice}
\end{figure}

\hfill \\[4pt]
\noindent
\textbf{{\large Results}}

\noindent
\textbf{Magnetism in DTN: brief introduction.} Magnetism in DTN originates from the spin-1 Ni$^{2+}$ ions forming a tetragonal lattice (we refer the reader to the Supplemental Material S1 for detailed crystallographic information). The magnetic properties are defined by the competition between the strong single-ion planar anisotropy $D$, and the antiferromagnetic exchanges interactions $J_{c}$ and $J_{a}\equiv J_{b}$ along the corresponding $c$ and $a,b$ directions. The effective Hamiltonian is:

\begin{equation}
 \label{EQ:DTNHamiltonian}
 \hamilt=\sum\limits_{\mathbf{r}}D(\spin^{z}_\textbf{r})^{2}+J_{c}\spop_{\mathbf{r}}\spop_{\mathbf{r+c}}+J_{a}\left(\spop_{\mathbf{r}}\spop_{\mathbf{r+a}}+\spop_{\mathbf{r}}\spop_{\mathbf{r+b}}\right).
\end{equation}
Here, $\mathbf{r}$ runs along the nickel positions in the tetragonal sublattice; $\mathbf{a,b}$, and $\mathbf{c}$ are the primitive lattice translation vectors towards the neighboring spins. Importantly, the symmetry prohibits any in-plane second-order anisotropic terms. The planar anisotropy protects the spin-singlet state $\ket{S^{z}=0}$ on every magnetic ion, thus preventing Ne\'el order in zero field, otherwise favored by the exchange interactions. The set of constants describing DTN at ambient pressure was initially obtained from zero-field neutron spectroscopy and thermodynamic studies~\cite{Zapf_PRL_2006_BECinDTN,ZapfCorrea_PRB_2008_DTNstriction}, and later refined as $D/\kb=8.9$~K, $J_{c}/\kb=1.82$~K, and $J_{a}/\kb=0.34$~K~\cite{PsaroudakiZvyagin_PRB_2012_DTNhighESR} based on the spectroscopic properties at high magnetic fields (where the effect of quantum renormalization is not present). The zero-field gap $\Delta/\kb\simeq3.5$~K~\cite{PaduanFilho_JAP_2004_DTNmagnetisation} is relatively small compared to the excitation-doublet bandwidth of  $\sim 8$~K. If the magnetic field is applied along the $c$ ($z$) direction, the degeneracy of the doublet is removed due to Zeeman splitting (as shown in Fig.~\ref{FIG:intro}b), triggering  eventually the onset of low-temperature magnetic order at $\mu_{0}H_{c1}\simeq2.1$~T~\cite{PaduanFilho_PRB_2004_DTNinfield,Zapf_PRL_2006_BECinDTN}. At stronger magnetic field, $\mu_{0}H_{c2}\simeq12.2$~T, DTN undergoes the transition
into the fully spin-polarized state.  We would like to stress, that the zero-field gap in DTN  is dominantly determined by the single-ion anisotropy (in contrast to  Haldane spin-chain materials), making DTN   a rare example of a large-$D$ spin-1 system.

\noindent
\textbf{High-pressure neutron diffraction.} As the pressure increases, the parameters of Hamiltonian Eq.~(\ref{EQ:DTNHamiltonian}) are expected to change. However, it is necessary to check first, whether the lattice remains undistorted. To this end, we have performed a series of structural neutron diffraction studies~\cite{Mannig_2017_PhDthesis}. As shown in Fig.~\ref{FIG:lattice}, the DTN lattice is smoothly compressible up to about $6$~kbar. Interestingly,  the compression goes in a rather uniaxial manner, with shrinking of the sample mostly along the $c$ axis (${\partial c}/{\partial P}=-0.038\pm0.003$~\AA/kbar at $1.8$~K). This implies the compression of the main Ni-Cl-Cl-Ni superexchange pathway and deformation of the Ni$^{2+}$ local environment. Such changes must affect $D$ and $J_c$ in the first place. The lattice parameter $a$ remains nearly constant with ${\partial a}/{\partial P}=(-4.1\pm1.5)\cdot10^{-3}$~\AA/kbar. At $P_\mathrm{irr}\simeq6$~kbar, an irreversible structural transition occurs, evidenced by discontinuities in the lattice parameters. More detailed discussion is given in the Supplemental Material (S2 and S3).

\noindent
\textbf{High-pressure TDO measurements.} The TDO susceptibility technique is well estalished as a versatile tool for detecting field-induced phase transitions in solids under applied pressure~\cite{ZvyaginGraf_NComm_2019_CCCpressure,ShiSteinhardt_NatComm_2019_dopedSCBO,ShiDissanayakeCorboz_NatComm_2022_SCBOpressureTDO,Coak_PRB_2023_PressuredDimers}.

\begin{figure}
  \includegraphics[width=0.5\textwidth]{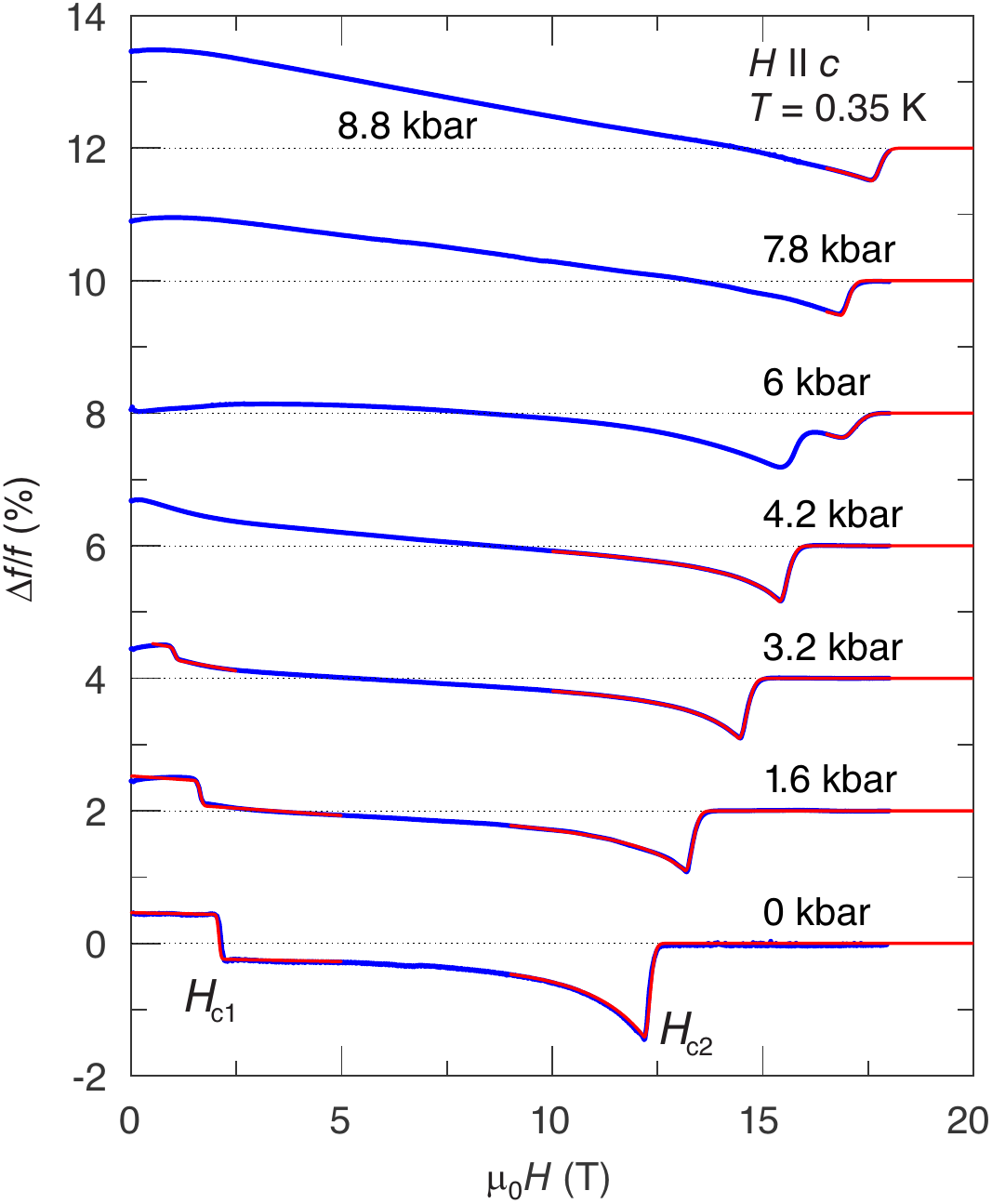}
  \caption{Pressure dependence of the TDO frequency changes $\Delta f/f$ in response to the magnetic ﬁeld (the data are
offset for clarity).  Results of empirical fits are shown in red (see Supplemental Material for details). }\label{FIG:TDO}
 \end{figure}

The results of our measurements are shown in Fig.~\ref{FIG:TDO}. At low pressures two anomalies, corresponding to the boundaries of the long-range AF ordered phase  at $H_{c1}$ and $H_{c2}$, respectively, are well discernible. To extract the critical fields in a reliable way we use a set of empirical functions (red lines, a detailed description is  given in the Supplemental Material). With applied pressure, we observe a decrease and increase of the first and second critical field, respectively. Above $4.2$~kbar, the low-field transition  is not visible anymore. This is a strong evidence of the spin-gap closure. The critical field $H_{c2}$ increases linearly up to $P_\mathrm{irr}$, where a discontinuity appears. The high-field part of the curve at $6$~kbar (about $P_\mathrm{irr}$) features a double-dip structure.  This reflects the coexistence of two structurally different phases at the first-order crystallographic transition.  In the structurally distorted phase above  $P_\mathrm{irr}$, we observe only one anomaly.

\noindent
\textbf{High-pressure ultrasound measurements.} In order to reveal the nature of the pressure-induced gapless phase, we performed high-pressure ultrasound measurements. The magnetic ordering in DTN at zero pressure was thoroughly investigated with ultrasound by Chiatti~\emph{et al.}~\cite{ChiattiSytcheva_PRB_2008_DTNsound,ChiattiZherlitsyn_JPConf_2009_DTNsound}, firmly establishing the connection between the long-range-order onset and the sound velocity anomalies in DTN. These anomalies reflects the spin-susceptibility divergence in the vicinity of the critical temperature~\cite{Luthi_2005_AcousticBook}, as a result of the pronounced magnetoelastic coupling in DTN.

\begin{figure}
  \includegraphics[width=0.5\textwidth]{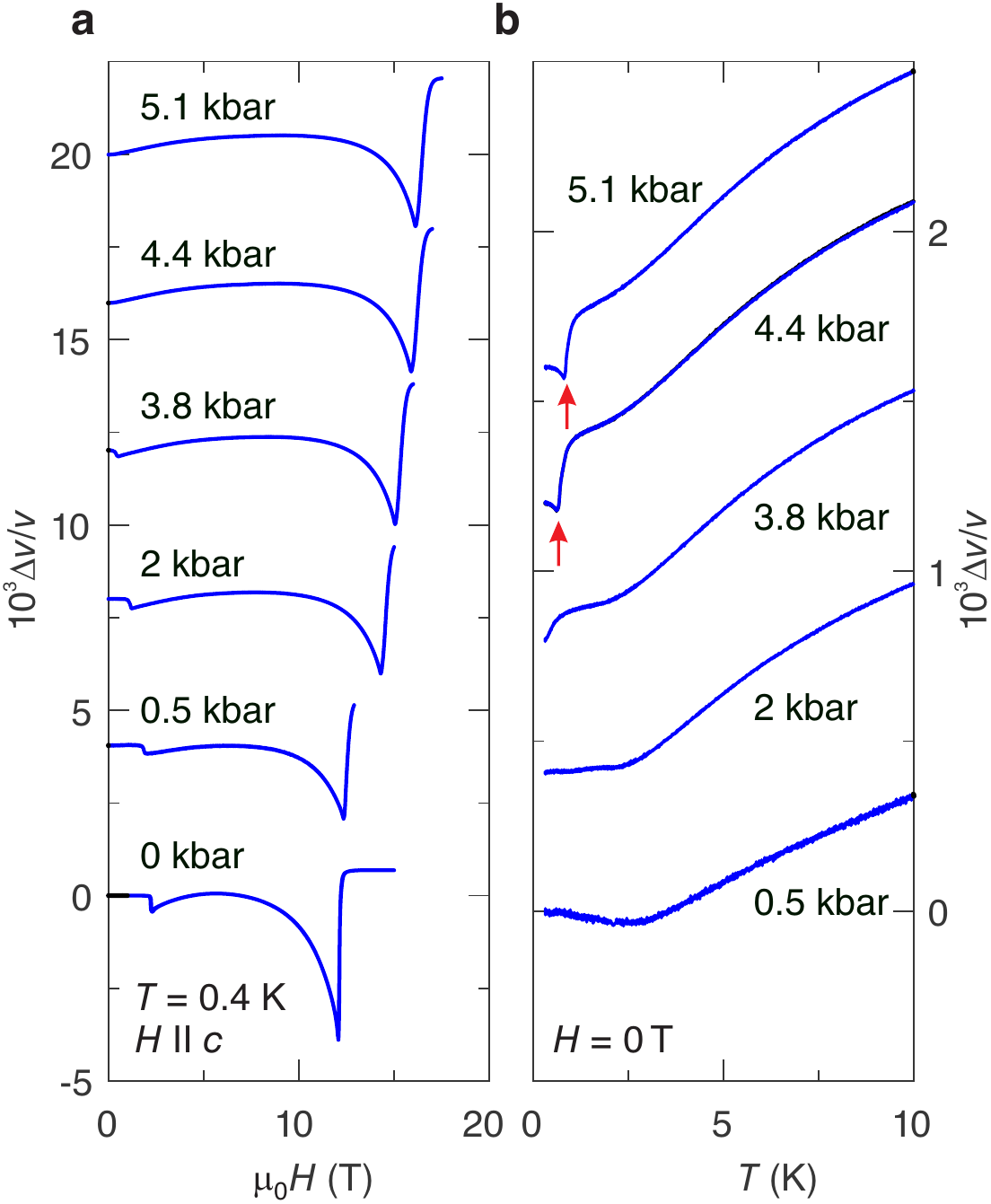}
  \caption{Pressure dependences of  ultrasonic properties of DTN.
     \textbf{a} Relative change of the sound velocity $\Delta v/v$ of the longitudinal acoustic mode  as function of magnetic field (zero-pressure data are taken  from Ref.~\cite{ChiattiSytcheva_PRB_2008_DTNsound} and were measured at $0.3$~K).  The data are offset for clarity. \textbf{b} Relative change of the sound  velocity  $\Delta v/v$ of the longitudinal acoustic mode as function of temperature. The data are offset for clarity.  The onset of magnetic order is marked by red arrows. }\label{FIG:C33}
 \end{figure}

We studied the longitudinal sound mode along the $c$ direction (mode $c_{33}$) at high pressures.
The field-induced changes of its velocity are shown in Fig.~\ref{FIG:C33}a.   Similar to our TDO data (Fig.~\ref{FIG:TDO}), there are two anomalies at low pressures. As revealed previously~\cite{ChiattiSytcheva_PRB_2008_DTNsound,ChiattiZherlitsyn_JPConf_2009_DTNsound}, they evidence the transitions from disordered gapped to AF ordered gapless state at $H_{c1}$, and the subsequent transition to the fully spin-polarized state at $H_{c2}$.

At higher pressures (above $3.8$~kbar), we only observe the high-field phase transition, while the gapless AF ordered phase becomes extended to zero magnetic field. An additional evidence for the pressure-induced magnetic ordering in zero field  is our observation of the sound-velocity anomaly   below $1$~K (red arrows in Fig.~\ref{FIG:C33}b). As expected,  the  ordering temperature grows with increasing pressure.

\begin{figure}
  \includegraphics[width=0.5\textwidth]{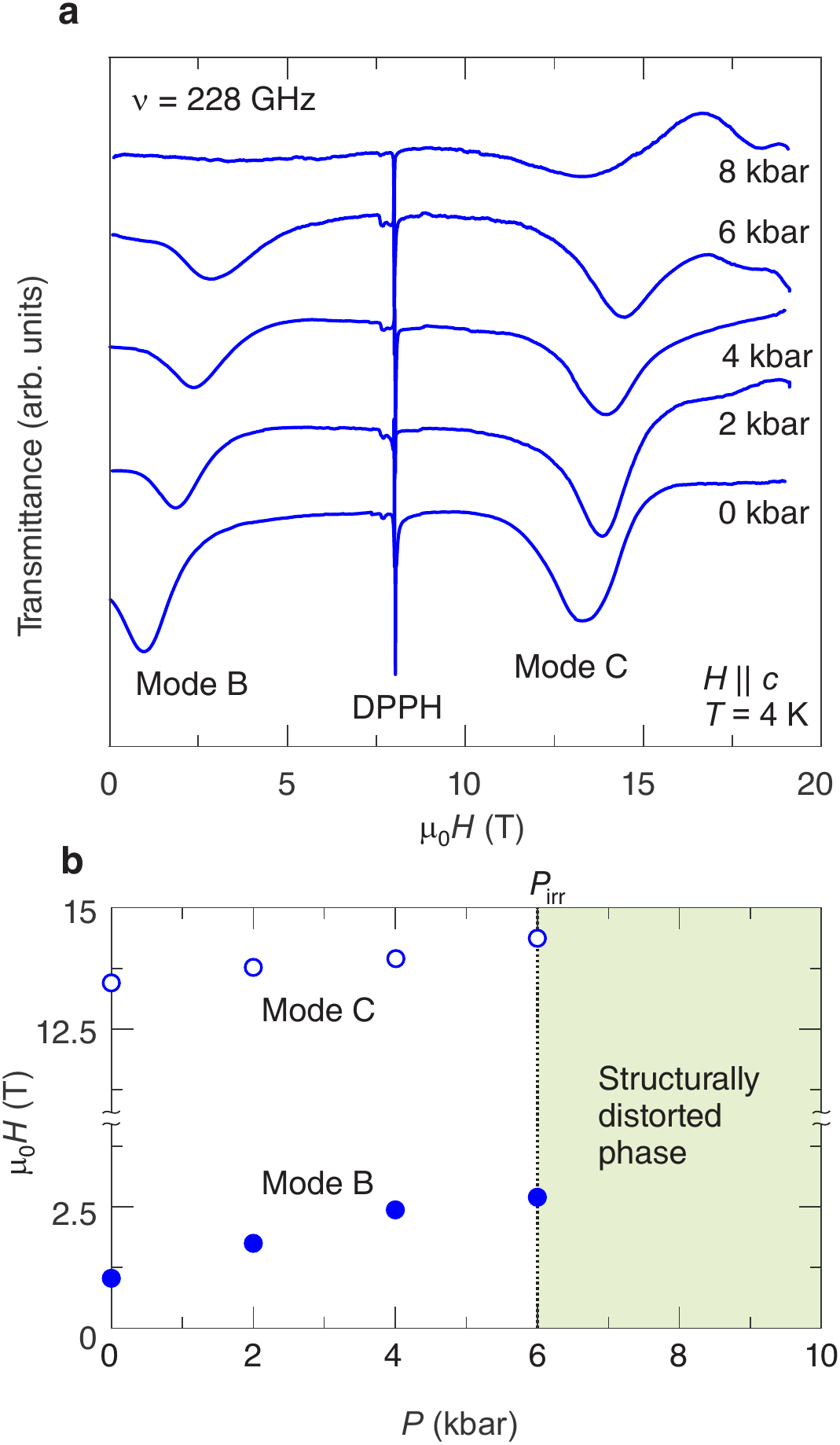}
  \caption{High-field ESR results. \textbf{a} Selected ESR spectra taken at a frequency of $228$ GHz ($H\parallel c$, $T=4$~K) at different  pressures [the sharp line corresponds to  DPPH (2,2-diphenyl-1-picrylhydrazyl), used as a marker]. \textbf{b} Pressure dependencies of ESR fields  for  mode B and C, as revealed by experiment.   }\label{FIG:ESR}
 \end{figure}

\noindent
\textbf{Electron spin resonance.} The applied pressure in DTN should not only affect the ground state but the  spin dynamics as well.  Electron spin resonance (ESR) has recently proven to be a powerful tool to probe the excitation spectra in strongly correlated spin systems under applied pressure ~\cite{SakuraiHirao_JPSJ_2018_SrCuBOpressureESR,ZvyaginGraf_NComm_2019_CCCpressure}. Selected examples of ESR spectra in DTN are shown in Fig.~\ref{FIG:ESR}a (adopting the naming convention from Ref.~\cite{PsaroudakiZvyagin_PRB_2012_DTNhighESR}, we label the observed modes B and C). Excitation  modes corresponding to $\Delta S^{z}=1$ transitions  remain well visible up to $6$~kbar. With the pressure increase both modes  demonstrate a gradual shift towards higher fields (as illustrated in Fig.~\ref{FIG:ESR}b) until they vanish in the  structurally distorted phase above about $6$~kbar. The complete frequency-field diagrams can be found in the Supplemental Material S6.

\hfill \\[4pt]
\noindent
\textbf{{\large Discussion}}

\begin{figure}
  \includegraphics[width=0.5\textwidth]{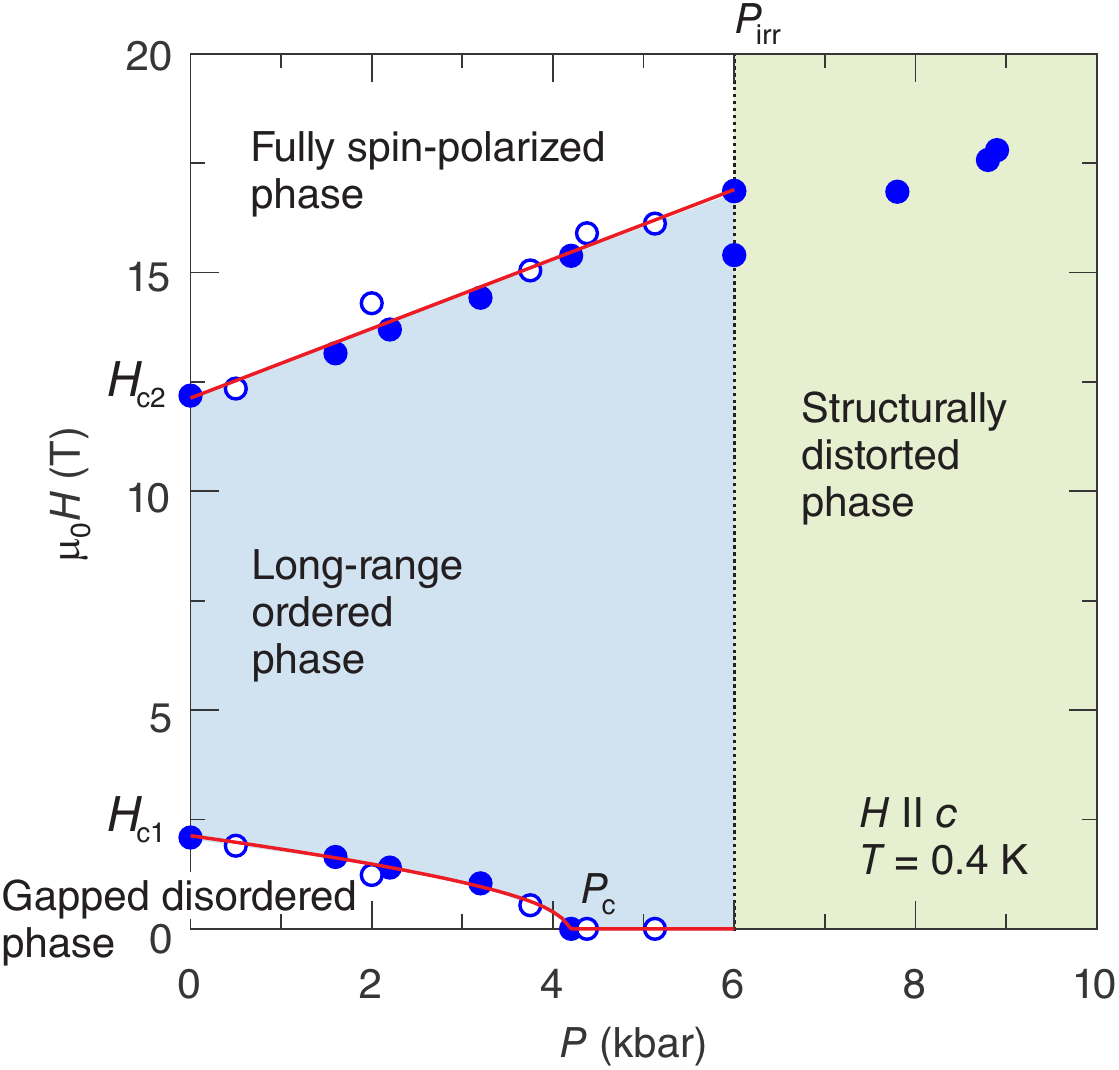}
  \caption{Field-pressure phase diagram of DTN. Open and closed symbols correspond to ultrasound and TDO data, respectively.
     Red lines are model fits (see the text for details). The dotted line corresponds to the irreversible structural phase transition at $P_\mathrm{irr} \sim 6$~kbar.}\label{FIG:phases}
 \end{figure}

The results of our TDO and ultrasound experiments on DTN are summarized in the phase diagram in Fig.~\ref{FIG:phases}. The data ensure that near $4$~kbar the spin gap does close and antiferromagnetic order emerges. One can theoretically describe the pressure-dependent spin Hamiltonian, consistent with these observations.  The saturation field $H_{c2}$ is known to follow the linear spin wave theory (LSWT) description without any quantum renormalization~\cite{PsaroudakiZvyagin_PRB_2012_DTNhighESR}:

\begin{equation}
 \label{EQ:Hc2}
 g\mub\mu_0 H_{c2}=D+4J_{c}+8J_{a}.
\end{equation}
Given the overall smallness of $J_{a}$ and lack of a significant pressure-induced length change in this direction, we can neglect possible variations in this interaction. Thus, the phase-diagram evolution is due to the changes in $D$ and $J_{c}$:

\begin{equation}
 \label{EQ:dHc2}
 g\mub\mu_0 \dfrac{\partial H_{c2}}{\partial P}=\dfrac{\partial D}{\partial P}+4\dfrac{\partial J_{c}}{\partial P}.
\end{equation}
With the $g$-factor being $2.26$, the measured slope $\mu_0{\partial H_{c2}}/{\partial P}=0.78\pm0.03$~T/kbar
(upper red line in Fig.~\ref{FIG:phases} below the instability pressure $P_\mathrm{irr}$) quantifies the linear pressure dependence of $D$ and $J_{c}$.

 \begin{figure}
  \includegraphics[width=0.5\textwidth]{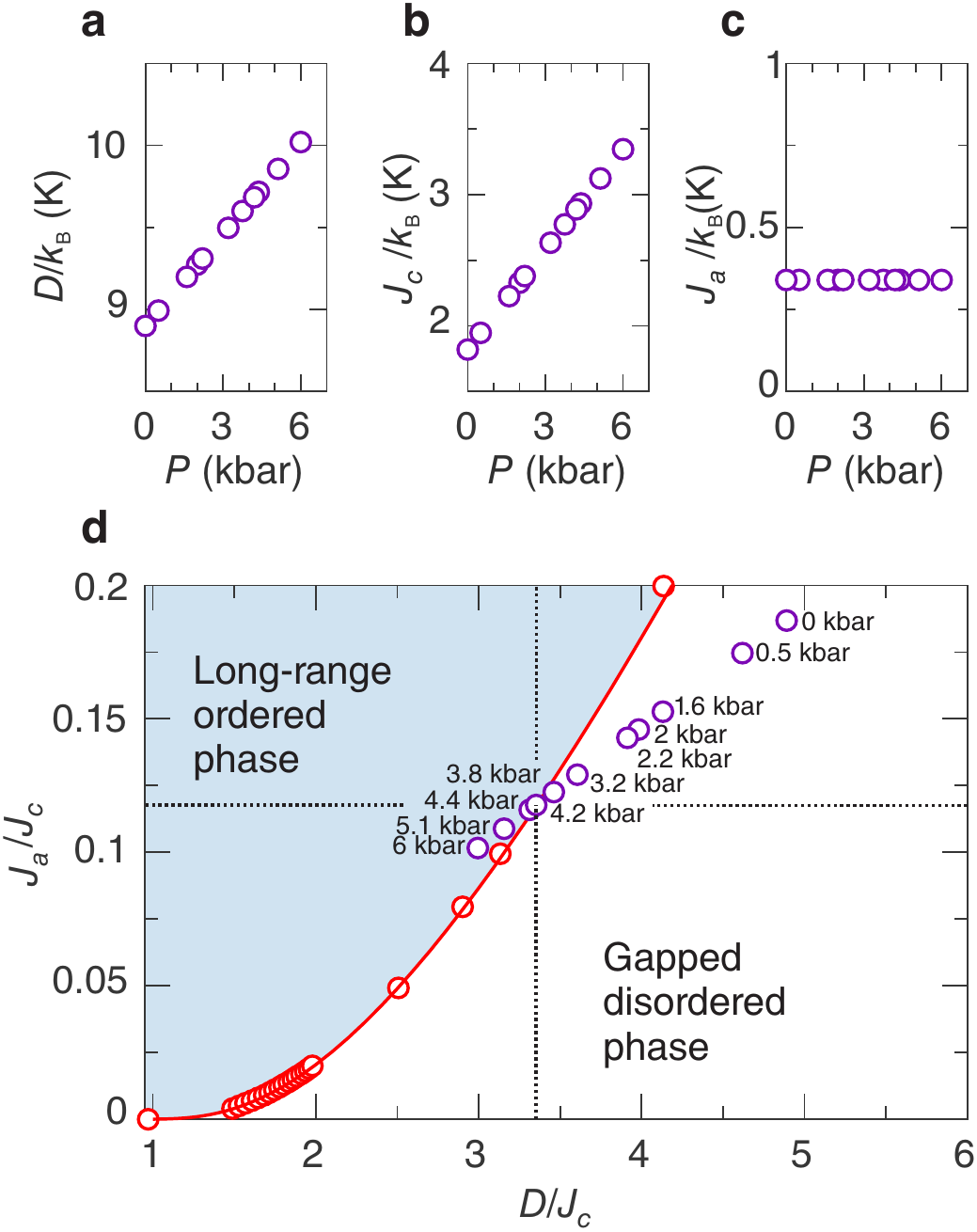}
  \caption{Results of model calculations. \textbf{a} Pressure dependence of the anisotropy parameter $D/\kb$. \textbf{b} Pressure dependence of the  exchange coupling  parameter $J_{c}/\kb$.  \textbf{c} Pressure dependence of the transverse exchange  coupling parameter $J_{a}/\kb$.   The calculation results are shown for the pressures, matching the experiments.
     \textbf{d} The Sakai--Takahashi  phase diagram for a quasi-1D spin-1 antiferromagnet~\cite{SakaiTakahashi_PRB_1990_DTNlikeGS}.  The red open points correspond to QMC calculations  for the phase boundary~\cite{WierschemSengupta_MPhysLettB_2014_DTNlikeGS}; the red line is a prediction based on the ansatz as described in the text. The purple circles correspond to  estimates of the spin Hamiltonian parameters at given pressures. The dotted cross marks the critical parameter values at $4.2$~kbar.}\label{FIG:ST}
 \end{figure}

The analysis of the first critical field poses a greater challenge: since $g\mu_{0}\mu_{B}H_{c1}=\Delta$, it requires evaluating the zero-field spin gap for the given set of Hamiltonian~(\ref{EQ:DTNHamiltonian}) parameters. No exact theory is available for that to the best of our knowledge. We overcome this challenge employing a random-phase approximation (RPA)~\cite{JensenMackintosh_1991_RareEarthBook} based ansatz combined with density matrix renormalization group (DMRG) calculations~\cite{White_PRL_1992_DMRGfirst,Schollwock_RMP_2005_DMRGreview}. The starting point is the determination of the zero-field gap value $\Delta_{0}(D/J_{c})$ in the $J_{a}=0$ limit of a single anisotropic chain, utilizing the latter technique. As far as we know, only a particular range close to the quantum critical point at $D/J_{c}\sim1$ was thoroughly accessed with DMRG earlier~\cite{Albuquerque_PRB_2009_HaldaneDgap}. The next step of the model is to apply the RPA treatment to interacting chains in order to estimate the critical value of the coupling $J_{a}^\mathrm{crit}$ that closes the gap. The details of the calculations (based on large-$D$ approximation for the dynamic structure factor~\cite{PapanicolaouSpatis_JPCM_1990_Spin1largeD,Lindgard_JMMM_1985_sumrules,*ZaliznyakLee_2005_NeutronSumRules} and the Kramers--Kronig relations) are given in the Supplemental Material (S4).

The main result is the approximation for the order-disorder zero-field phase boundary:

\begin{equation}\label{EQ:CritCoupling}
 J_{a}^\mathrm{crit}=\dfrac{\Delta_{0}^2}{4\mathcal{A}D}.
\end{equation}
We find that the numerical value $1/4\mathcal{A}\simeq0.14$ provides an excellent description of this phase boundary, previously obtained by extensive quantum Monte--Carlo (QMC) simulations~\cite{SakaiTakahashi_PRB_1990_DTNlikeGS,WierschemSengupta_MPhysLettB_2014_DTNlikeGS}. This comparison can be seen in Fig.~\ref{FIG:ST}d, where the calculation result using Eq.~(\ref{EQ:CritCoupling}) is shown as the red line. Then, it follows from our approach, that the value of the gap, renormalized by $J_{a}$, is

\begin{equation}\label{EQ:GapAnsatz}
 \Delta=\Delta_{0}\sqrt{1-\dfrac{J_{a}}{J_{a}^\mathrm{crit}}}.
\end{equation}
Hence, the first critical field can be expressed through the pressure-dependent Hamiltonian parameters as

\begin{equation}\label{EQ:AnsatzHc1}
 g\mu_{0}\mu_{B}H_{c1}(P)=\Delta\left[\Delta_{0}\left(\frac{D(P)}{J_{c}(P)}\right), D(P)\right].
 \end{equation}
The ambient pressure $H_{c1}$ evaluated by this procedure agrees with the experimental value, ensuring the validity of the approach for the smaller gap values as well.

 The ability to predict both critical fields using Eqs.~(\ref{EQ:Hc2}) and~(\ref{EQ:AnsatzHc1}) finally opens a route to a self-consistent treatment of the measured phase diagram without any ``effective'' parameters. We numerically optimize the agreement between~(\ref{EQ:AnsatzHc1}) and the experimental $H_{c1}$ data by varying the choice of the critical pressure $P_{c}$ at which the gap closes completely. This fixes a unique combination of $\partial (D, J_{c})/\partial P$. The optimization results in the critical pressure $P_{c}=4.2\pm0.3$~kbar (see Supplemental Material S5 for more details). The magnetic phase diagram (Fig.~\ref{FIG:phases}) is fully captured with $\kb^{-1}\partial D/\partial P=0.16\pm0.03$~K/kbar and $\kb^{-1}\partial J_{c}/\partial P=0.25\pm0.01$~K/kbar. Figs.~\ref{FIG:ST}a-c illustrate how the spin-Hamiltonian parameters of DTN are pressure-tuned up to $P_\mathrm{irr}$. In the phase diagram of Fig.~\ref{FIG:ST}d, we show how the corresponding ground state (illustrated by purple points) is changing from gapped to long-range ordered at $P_c$ in accordance with the previous discussion.

The rich high-pressure physics of DTN under pressure invites further investigations, related to the interplay of quantum and thermal fluctuations near $z=1$ quantum criticality~\cite{Merchant_NatPhys_2014_PindTlCuCl,ScammellSushkov_PRB_2015_AsympFreedom,*ScammellSushkov_PRB_2017_Paramagnons,ScammellSushkov_PRB_2017_Multiuniversal,HongQiu_NatComm_2017_DLCBrenorms}. In particular, it would be important to access the order parameter (e.g., with neutron diffraction or nuclear magnetic resonance) and the extended field-temperature-pressure phase diagram~\cite{Coak_PRB_2023_PressuredDimers}. This would open the possibility to establish a direct connection between the minimalistic and highly symmetric spin Hamiltonian of DTN and the effective field theory used to describe the static and dynamic properties of critical quantum magnets~\cite{ScammellKharkov_PRB_2017_O3critical,QinNormand_PRL_2017_HiggsQMC}. The undistorted symmetry also makes DTN a perfect candidate for investigating of the effect of thermal and quantum fluctuations on the amplitude mode~\cite{Merchant_NatPhys_2014_PindTlCuCl} in the absence of Ising-type anisotropy.
One can also expect chemically substituted DTNX~\cite{PovarovMannig_PRB_2017_DTNXcrit,*MannigPovarov_PRB_2018_DTNXGSWT} to be a fruitful playground for the interplay of quantum $z=1$ criticality and quenched disorder, similarly to (C$_4$H$_{12}$N$_2$)Cu$_2$(Cl$_{1-x}$Br$_x$)$_6$~\cite{MannigMoeller_PRB_2016_PHCXmuons}, (NH$_4$)$_x$K$_{1-x}$CuCl$_3$~\cite{KinyonDalal_PRM_2021_XKCuCl3}, and \XFeCl~\cite{HayashidaStoppel_PRB_2019_CsFeCl3disorderINS}. The higher symmetry and a simpler, well-understood Hamiltonian make the case of DTNX much more appealing for such studies.

To summarize, we have identified the material DTN as a unique platform for studies of the exotic $z=1$ universality class in a three-dimensional magnetic material. By means of TDO, ultrasound, and ESR measurements we have confirmed the existence of pressure-induced criticality, separating gapped disordered and gapless long-range ordered magnetic phases in the material. The nearly ideal axial symmetry of the structure is retained at that point, which makes DTN to stand out among the non-frustrated quantum magnets. In addition to that, we have extracted the pressure dependence of the spin Hamiltonian parameters from the data, and have achieved a quantitative theoretical understanding of the transition mechanism.

\hfill \\[4pt]
\noindent
\textbf{{\large Methods}}

\noindent
\textbf{Sample growth.} Samples for the thermodynamic measurements and the electron spin resonance spectroscopy were synthesized in the University of Sa\~o Paulo  from aqueous solution using a thermal-gradient method~\cite{PaduanFilho_JChemPhys_1981_DTNfirst,*Yankova_PhilMag_2012_ReviewXtals}. Samples used in the neutron diffraction studies were synthesized at ETH Z\"{u}rich. Fully deuterated chemicals (water and thiourea) were used. The crystals were crushed into powder for these experiments.

\noindent
\textbf{High-pressure neutron diffraction.} Neutron-diffraction experiments with the powder samples were performed at instrument HB2a at the High Flux Isotope Reactor, Oak Ridge National Laboratory (Oak Ridge, Tennessee, USA). Up to a few grams of deuterated DTN powder material were used. A Ge[113] or Ge[115] vertically focussing wafer-stack monochromator was used to produce a neutron beam with $2.41$~\AA\ or $1.54$~\AA\ wavelength, respectively. A $^4$He Orange cryostat with aluminum He-gas pressure cell (capable of producing pressures up to $6$~kbar) and a CuBe clamp cell (capable of producing pressures up to $20$~kbar) were used. The pressure in the gas cell was measured in situ using a manometer. For the clamp cell a rock salt (halite) calibration curve was used for the pressure estimate, resulting in an error bar of about $1$~kbar. After initial data reduction, the \texttt{FULLPROF} package~\cite{RodriguezCarvjal_PhysicaB_1993_FULLPROF} was used for the intensity profile analysis and structure determination. More details of these experiments can be found in Ref.~\cite{Mannig_2017_PhDthesis}.

\noindent
\textbf{High-pressure TDO.} High-pressure TDO measurements were conducted at the National High Magnetic Field Laboratory, Florida State University (Tallahassee, Florida, USA) in magnetic fields up to 18 T using a TDO susceptometer~\cite{CloverWulf_RevSciInstr_1970_TDO}. Magnetic field was applied along the $c$ axis of the crystal placed in a tiny copper-wire coil with $0.8$~mm diameter and $1$~mm height. This assembly was immersed into Daphne 7575 oil and encapsulated in a Teflon cup inside the bore of a piston-cylinder pressure cell made of a chromium alloy (MP35N). The pressure created in the cell was calibrated at room temperature and again at low temperature using the fluorescence of the $R1$ peak of a small ruby chip as a pressure marker~\cite{Piermarini_JAplPhys_1975_R1ruby} with accuracy better than $0.15$~kbar. The pressure cell was immersed directly into $^3$He, allowing TDO measurements down to $350$~mK.

\noindent
\textbf{High-pressure ultrasound measurements.} We performed ultrasound measurements using the pulse-echo method with phase-sensitive detection technique~\cite{WolfLuthi_PhysB_2001_UltrasoundReview, Luthi_2005_AcousticBook}. Overtone polished LiNbO$_3$ transducers with 36${^\circ}$ Y-cut (longitudinal sound polarization) were used to generate and detect ultrasonic signals with a frequency of about $35$~MHz. The transducers were bonded to natural $(001)$ crystal surfaces of a DTN sample with Thiokol 32, providing $k\parallel u \parallel H \parallel c$ experiment geometry. The crystal dimensions were \mbox{$ 0.3 \times 0.3 \times 2.95 $ mm$^3$}.  We determined the low-$T$ sound velocity at zero field and pressure to $v \simeq 2600 $~m/s, in agreement with~\cite{ChiattiSytcheva_PRB_2008_DTNsound,ChiattiZherlitsyn_JPConf_2009_DTNsound}. A commercially available piston-cylinder cell with CuBe outer sleeve, NiCrAl inner sleeve, and tungsten carbide inner pistons (C\&T Factory Co.,Ltd) was adapted for ultrasound experiments following ~\cite{Mombetsu_PRB_2016_UltrasoundPressure}, with Daphne oil 7373 as a pressure medium. The pressure cell (with a calibrated RuO$_2$ temperature sensor on the outer side) was thermally anchored to the $^3$He pot of the cryostat via a copper rod. The pressure was estimated from the pressure-dependent superconducting transition of tin~\cite{SmithWu_PR_1967_TinSCpressure,*EilingSchilling_JPhysF_1981_SimpleSCpressure}.

\noindent
\textbf{High-pressure ESR.}
High-pressure ESR studies of DTN were performed employing a $25$~T cryogen-free superconducting magnet ESR setup (25T-CSM) at the High Field Laboratory for Superconducting Materials, Institute for Materials Research, Tohoku University (Sendai, Japan)~\cite{Sakurai_ApplMagRes_2021_PressESR,SakuraiKimura_JMRes_2018_THzPress25T}. Gunn diodes were utilized as microwave sources for frequencies up to $405$~GHz; the transmitted radiation power was detected using a hot-electron InSb bolometer operated at $4.2$~K. A DTN crystal was loaded into a Teflon cup filled with Daphne 7474 oil as a pressure medium. A two-section piston-cylinder pressure cell made from NiCrAl (inner cylinder) and CuBe (outer sleeve) has been used. The key feature of the pressure cell is the inner pistons, made of ZrO$_2$ ceramics. The applied pressure was calibrated against the superconducting transition temperature of  tin~\cite{SmithWu_PR_1967_TinSCpressure,*EilingSchilling_JPhysF_1981_SimpleSCpressure}, detected by AC susceptometer. The actual pressure during the experiment was calculated using the relation between the load at room temperature and the pressure obtained at around $3$ K; the pressure calibration accuracy was better than $0.5$~kbar~\cite{SakuraiFujimoto_JMagnRes_2015_pressreESR}.

\noindent
\textbf{DMRG calculations.}
The DMRG calculations utilized the  \texttt{Julia} version of the \texttt{ITensors} package~\cite{Bezanson_SIAM_2017_Julia,FishmanWhite_SciPost_2022_ITensors}. A $249$-site anisotropic $S=1$ chain was simulated. The calculations were performed on the \texttt{hemera} cluster (HZDR).

\hfill \\[4pt]
\noindent
\textbf{Data availability}
The data that support the findings of this study are available from the corresponding author upon request.

\bibliography{DTNTDOlib_v22}

\noindent
\textbf{Acknowledgements}
We thank A. Mannig, J. M\"oller, and G. Perren for their involvement at the early stage of the ETH Z\"{u}rich part of the project. This work was supported by the Deutsche Forschungsgemeinschaft through ZV 6/2-2, the W\"{u}rzburg-Dresden Cluster of Excellence on Complexity and Topology in Quantum Matter - $ct.qmat$ (EXC 2147, project No. 390858490) and the SFB 1143 (project No. 247310070), as well as by HLD at HZDR,  member of the European Magnetic Field Laboratory (EMFL) [K.Yu.P., S.A.Zv., S.Zh., A.H., J.W.].
 The neutron-diffraction experiments at HB2a used resources at the High Flux Isotope Reactor, a Department of Energy Office of Science User Facility operated by the Oak Ridge National Laboratory. This work was partially supported by the Swiss National Science Foundation, Division II [A.Z.].  A portion of this work has been performed at the National High Magnetic Field Laboratory, which is supported by the National Science Foundation Cooperative Agreement DMR-1644779 and the State of Florida [D.E.G.]. ESR experiments were performed at the High Field Laboratory for Superconducting Materials, Institute for Materials Research, Tohoku University (proposal 19H0501 and 20H0501). Support of the ICC-IMR Visitor Program at Tohoku University is acknowledged [S.A.Zv.]. This work was partially supported  by the Brazilian agencies CNPq (grant 304455-2021-0) and FAPESP (grant 2021-12470-8) [A.P.F.]. We also thank M. E. Zhitomirsky for fruitful discussions.

\hfill \break

\noindent
\textbf{Author contributions}
S.A.Zv. conceived the project and performed the TDO (together with D.E.G.) and ESR (together with T.S., S.K., and H.N.) measurements; K.Yu.P. performed the analysis of this data and the numeric simulations.
 V.O.G. has performed the neutron diffraction measurements and analyzed the data. A.H. and S.Zh. have performed the ultrasound propagation measurements. M.N. has contributed to the preliminary high-pressure thermodynamic measurements. A.P.F. has synthesized the non-deuterated single crystals used in the study. H.O. and H.N., A.Z., and J.W. administered at the work at Kobe University and Tohoku University, ETH Z\"{u}rich, and HZDR, respectively. All the authors have contributed to the discussions and manuscript writing.

  \noindent
\textbf{Competing interests}
The authors declare no competing interests.

 

\section*{Supplementary material}

# Supplemental Material for the manuscript
## "Pressure-tuned quantum criticality in the large-$D$ antiferromagnet DTN"

Kirill Yu. Povarov,[1, *] David E. Graf,[2] Andreas Hauspurg,[1, 3] Sergei Zherlitsyn,[1] Joachim Wosnitza,[1, 3] Takahiro Sakurai,[4] Hitoshi Ohta,[5, 6] Shojiro Kimura,[7] Hiroyuki Nojiri,[7] V. Ovidiu Garlea,[8] Andrey Zheludev,[9] Armando Paduan-Filho,[10] Michael Nicklas,[11] and Sergei A. Zvyagin[1, †]

[1]*Dresden High Magnetic Field Laboratory (HLD-EMFL) and Würzburg-Dresden Cluster of Excellence ct.qmat, Helmholtz-Zentrum Dresden-Rossendorf (HZDR), 01328 Dresden, Germany*
[2]*National High Magnetic Field Laboratory, Tallahassee, Florida 32310, USA*
[3]*Institut für Festkörper- und Materialphysik, Technische Universität Dresden, 01062 Dresden, Germany*
[4]*Research Facility Center for Science and Technology, Kobe University, Kobe, 657-8501, Japan*
[5]*Molecular Photoscience Research Center, Kobe University, Kobe 657-8501, Japan*
[6]*Graduate School of Science, Kobe University, Kobe 657-8501, Japan*
[7]*Institute for Materials Research, Tohoku University, Sendai 980-8578, Japan*
[8]*Neutron Scattering Division, Oak Ridge National Laboratory, Oak Ridge, TN 37831, USA*
[9]*Laboratory for Solid State Physics, ETH Zürich, Switzerland*
[10]*Instituto de Fisica, Universidade de São Paulo, 05315-970 São Paulo, Brazil*
[11]*Max Planck Institute for Chemical Physics of Solids, 01187 Dresden, Germany*
(Dated: February 23, 2024)

This Supplemental Material contains crystallographic information, details of neutron-diffraction experiments under pressure, estimates of lattice compressibility, the numerical/analytical ansatz used to describe the first critical field, various fit procedures, and additional high-pressure ESR measurements.

## CONTENTS

---

* k.povarov@hzdr.de; Previous address: Laboratory for Solid State Physics, ETH Zürich, Switzerland
† s.zvyagin@hzdr.de

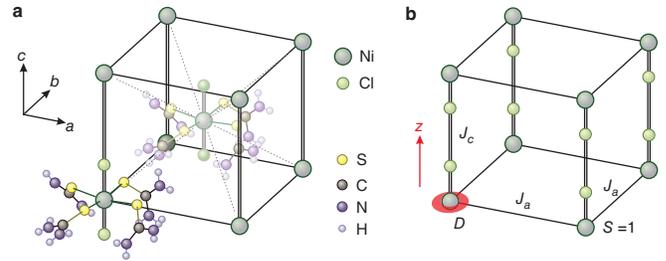

FIG. S1. **a** Crystal structure of DTN, with two NiCl$_2 \cdot$4SC(NH$_2$)$_2$ units per body-centered tetragonal cell. **b** Sketch of the relevant magnetic interactions in DTN. The single-ion easy-plane perpendicular to $c$ ($z$) direction is highlighted for one of the Ni$^{2+}$ ions.

## I. CRYSTALLOGRAPHIC INFORMATION

A schematic view of DTN crystallographic structure (body-centered tetragonal, space group $I4$) is displayed in Fig. S1a. The magnetic properties are determined by the competition between the strong single-ion planar anisotropy $D$ and the antiferromagnetic exchange couplings $J_c$ (along the $c$ direction, Ni-Cl-Cl-Ni superexchange bond) and $J_a$ along the $a, b$ directions. This is illustrated in Fig. S1b. Possible interactions between the two interpenetrating tetragonal sublattices are known to be frustrated and negligible [1].

## II. HIGH-PRESSURE NEUTRON-DIFFRACTION EXPERIMENTS

Neutron-diffraction experiments on powder samples were performed using instrument HB2a, at the High Flux Isotope Reactor, Oak Ridge National Laboratory (Tennessee, USA). Up to a few grams of deuterated DTN powder material were used. A Ge[113] or Ge[115] vertically focussing wafer-stack monochromator was used to produce a neutron beam with 2.41 Å or 1.54 Å wavelength. A $^4$He Orange cryostat with aluminum He-gas pressure cell (capable to produce pressures up to 6 kbar) or a CuBe clamp cell were used.

The pressure in the gas cell was measured *in situ* using a manometer. For the clamp cell a rock salt calibration curve was utilized, that provided accuracy better than 1 kbar.

Selected diffraction datasets (scattered neutron beam intensity vs momentum transfer) are shown in Fig. S2. After the initial data reduction, the FULLPROF package [2] was used for the intensity-profile analysis and structure determination. Fit examples are shown in Fig. S3. At small pressures, the peak pattern remains qualitatively the same for all temperatures measured, only the lattice constants are decreasing gradually. The situation changes at around $P_{\rm irr} \simeq 6$ kbar. Extra Bragg peaks appear, revealing a doubling of the structural unit cell. These peaks are not vanishing upon releasing the pressure (Fig. S2b), evidencing the irreversible character of the phase transition. A more detailed analysis of this transition and the high-pressure structure is given in Ref. [3].

## III. FITS TO EXTRACT THE CRITICAL FIELDS

Here, we discuss the fits describing the tunnel-diode-oscillator (TDO) frequency-shift data (Fig. 3 of the main text). For the low-field step-like anomaly, we use the following empirical fit function:

$$\frac{\Delta f}{f} = \left[\frac{\Delta f}{f}\right]_1 + \eta_1 H + \frac{A_1}{2}\left\{1 + \mathrm{erf}\left(\frac{H - H_{c1}}{\delta H_1 \sqrt{2}}\right)\right\}. \tag{S.1}$$

Here, the first two terms describe the linear background, and the step-like part is described by the Gaussian error function centered at the first critical field $H_{c1}$. The parameter $\delta H_1$ controls the transition width.

Describing the TDO data at $H_{c2}$ is more complicated, as one has to deal with an asymmetric peak there. We use the following empirical formula:

$$\frac{\Delta f}{f} = \left[\frac{\Delta f}{f}\right]_2 + \eta_2 H + \begin{cases} \dfrac{A_2}{\delta H_2 \sqrt{2\pi}} \exp\left[-\dfrac{1}{2}\left(\dfrac{H - H_{c2}}{\delta H_2}\right)^2\right], & H \geq H_{c2} \\ B + \dfrac{A'}{(H_\infty - H)^\alpha}, & H < H_{c2} \end{cases}. \tag{S.2}$$

Here, the high-field part of the peak is described by a Gaussian distribution of weight $A_2$ and width $\delta H_2$, centered at $H_{c2}$. Below this critical field, the broad part is better captured by a generalized hyperbola, centered at some effective field $H_\infty > H_{c2}$, with the amplitude $A'$ and the characteristic exponent $\alpha$. The offset $B = \dfrac{A_2}{\delta H_2 \sqrt{2\pi}} - \dfrac{A'}{(H_\infty - H_{c2})^\alpha}$ ensures that the curve is continuous at $H_{c2}$. The linear background, characterized by $\left[\dfrac{\Delta f}{f}\right]_2$ and $\eta_2$, is the same for both parts of the curve. This background is subtracted from the full range of the data in order to yield the detrended version of the plots actually shown in Fig. 3.

## IV. ESTIMATING THE SPIN GAP MAGNITUDE

### A. DMRG results

As mentioned in the main text, we simulated an $N = 249$ sites $S = 1$ chain, using the density matrix renormalization group (DMRG) implementation in the Julia version of ITensors package (30 sweeps, energy convergence better than $10^{-6}$). In particular, it allows for a straightforward gap-size estimation. First, one finds the ground-state vector in the Hilbert space of the model, which is a conventional DMRG routine. Then, this ground state is excluded from the Hilbert space, and the DMRG procedure is repeated once again. Naturally, the lowest energy state yielded is then the first excited state. The energy difference between the two respective states corresponds to the energy gap $\Delta_0$. The calculation results are shown in Fig. S4. They provide a very good interpolation between the two known limits of the model. On the large-$D$ end the generalized spin-wave

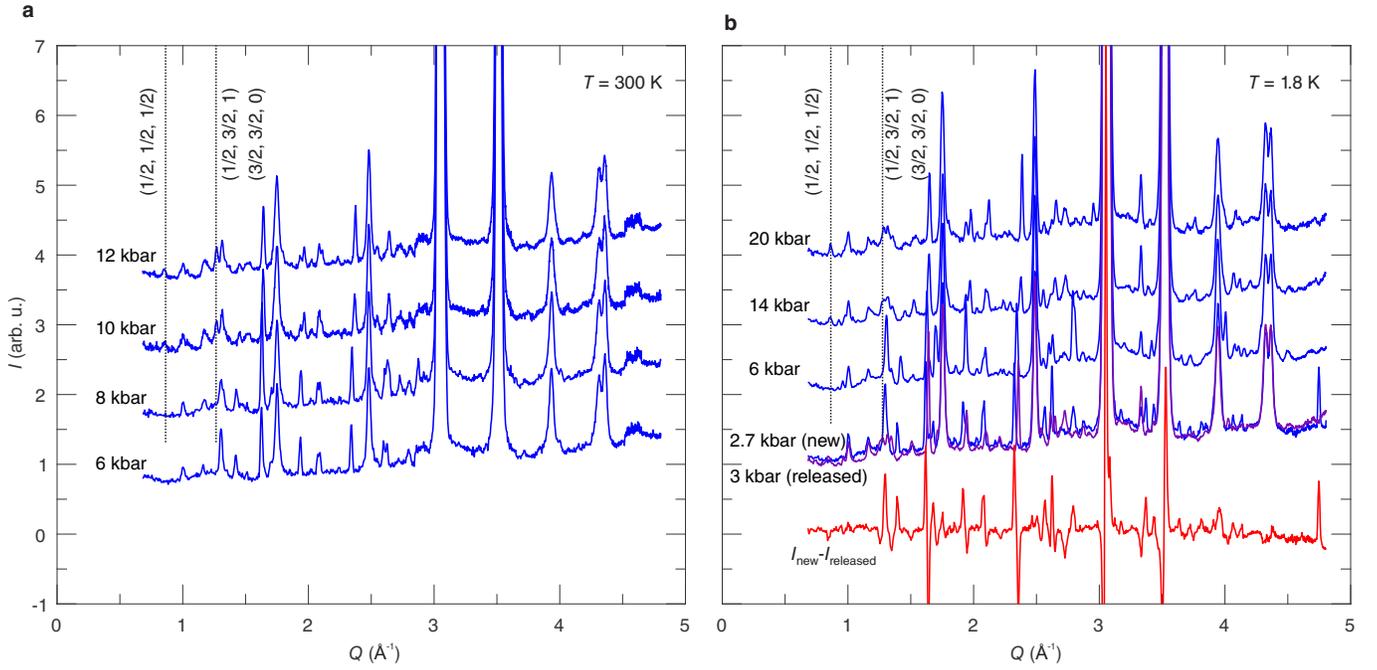

FIG. S2. Measured neutron-diffraction intensity vs momentum transfer at different pressures [3]. **a** Data at 300 K, clamp cell. The data is offset for clarity. The dotted lines indicate the position of some Bragg peaks that correspond to the lattice period doubling. **b** Data at 1.8 K, clamp cell. The 3 kbar curve obtained after the pressure release (purple), is overlayed with the 2.7 kbar curve obtained at the initial loading cycle (blue). The intensity difference between these two measurements is shown below in red.

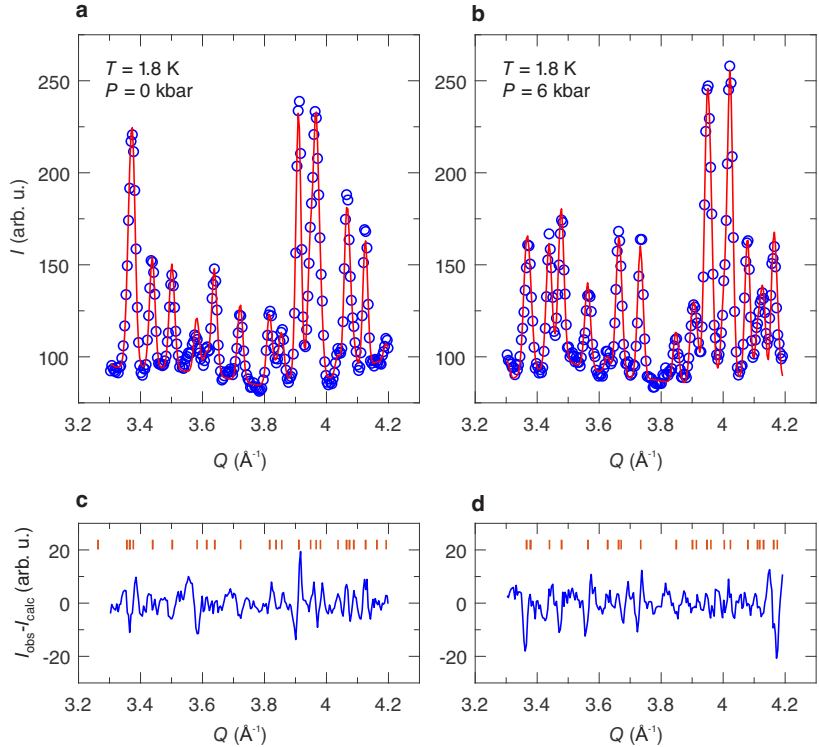

FIG. S3. Examples of fits (red lines) to the powder diffraction data (open circles) of DTN [3]. **a** Data at ambient pressure. **b** Data at 6 kbar. **c,d** The residues and the expected Bragg peak positions.

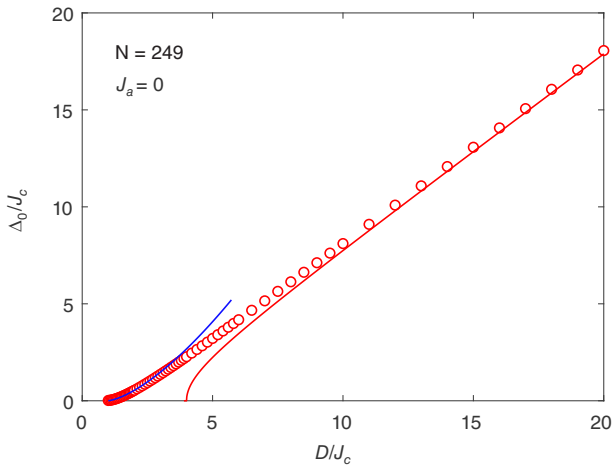

FIG. S4. The gap $\Delta_0/J_c$ in the disordered large-$D$ phase for the one-dimensional model as function of $D/J_c$. The points are the results of DMRG calculations. The red solid line represents GSWT results in the large-$D$ limit. The blue line shows the critical gap scaling near $D/J_c \simeq 1$.

theory (GSWT) is applicable. It describes the spin-gap as $\sqrt{D^2 - 4DJ_c}/J_c$ [4–6]. On the other end, in the vicinity of the critical point one finds the spin-gap critical behavior $0.52(D/J_c - 0.971)^{1.478}$ [7]. These two regimes are shown as red and blue lines in Fig. S4 correspondingly.

### B. Critical coupling from RPA

Here, we give a detailed description of the random phase approximation (RPA)-based *ansatz*, which we used to calculate the gap (and, hence, $H_{c1}$) as a function of pressure. The starting point of our calculation is the RPA (or a mean-field, if one prefers such notation) expression for the transverse staggered susceptibility [8]:

$$\chi_{xx}^{\prime \mathrm{RPA}}(\mathbf{Q}) = \frac{\chi_{xx}^{\prime \mathrm{1D}}(\pi/c)}{1 - 4J_a \chi_{xx}^{\prime \mathrm{1D}}(\pi/c)}. \quad (\mathrm{S}.3)$$

Here, the three-dimensional antiferromagnetic propagation vector is $\mathbf{Q} = (\pi/a,\ \pi/a,\ \pi/c)$, and $\chi_{xx}^{\prime \mathrm{1D}}(\pi/c)$ is the transverse staggered susceptibility of a single chain. Eq. (S.3) describes the critical susceptibility that diverges once we enter the domain of planar long-range order.

To evaluate this quantity, we employ certain dynamic structure-factor approximations for large $D$ and the Kramers–Kronig relations. The useful RPA-type approximation we make for the transverse dynamic structure factor in a large-$D$ chain of $N$ sites is:

$$\mathcal{S}^{xx}(\pi/c, \omega) = N \frac{\mathcal{A}D}{\Delta_0} \delta(\Delta_0 - \omega). \quad (\mathrm{S}.4)$$

The dispersion relation has a minimum at the wavevector $\pi/c$. The excitation energy at this minimum is the gap $\Delta_0$. The amplitude prefactor $\mathcal{A}$ is, generally, a complex quantity that depends on $D$ and $J_c$, as well as on the on-site and inter-site ground-state correlators $\left\langle (\hat{S}^x)^2 \right\rangle$ and $\left\langle \hat{S}_n^\alpha \hat{S}_{n+1}^\alpha \right\rangle$, with $\alpha = x, z$ [9, 10]. However, in first approximation this is merely $\mathcal{A} = 1 + \mathcal{O}(J_c/D)$. Treating $\mathcal{A}$ as a constant somewhat larger than 1 is the approximation which we adopted in our *ansatz* beyond conventional RPA.

Then, at $T = 0$ the dissipative part of the transverse staggered susceptibility has the following form according to the fluctuation-dissipation theorem:

$$\chi_{xx}^{\prime\prime}(\pi/c, \omega) = \frac{\pi}{N}\mathcal{S}^{xx}(\pi/c, \omega) = \frac{\pi \mathcal{A} D}{\Delta_0}\delta(\Delta_0 - \omega). \quad (\mathrm{S}.5)$$

Now, using the Kramers–Kronig relations, one can evaluate the desired static susceptibility, which is equivalent to the reactive part at $\omega = 0$:

$$\chi_{xx}^{\prime}(\pi/c, 0) = \frac{1}{\pi} \int\limits_{-\infty}^{+\infty} \frac{\pi \mathcal{A} D}{\Delta_0} \frac{\delta(\Delta_0 - \Omega)}{\Omega} d\Omega \quad (\mathrm{S}.6)$$

We now have obtained the following estimate for the one-dimensional critical susceptibility:

$$\chi_{xx}^{\prime}(\pi/c) = \frac{\mathcal{A}D}{\Delta_0^2}, \quad (\mathrm{S}.7)$$

which, combined with the RPA expression (S.3) results in the ordering criterion:

$$4J_a \frac{\mathcal{A}D}{\Delta_0^2} = 1. \quad (\mathrm{S}.8)$$

This is the criterion (5) from the main text.

### C. The gap in three dimensions

Now, we can estimate the influence of $J_a$ bonds on the spin gap $\Delta_0$. This gap value, modified by the three-dimensional interactions, we label $\Delta_{3\mathrm{D}}$. Using this notation, we can write a three-dimensional analogue of Eq. (S.7):

$$\chi_{xx}^{\prime \mathrm{RPA}}(\mathbf{Q}) = \frac{\mathcal{A}D}{\Delta_{3\mathrm{D}}^2}, \quad (\mathrm{S}.9)$$

where we assume a constant factor $\mathcal{A}$. Then, using Eqs. (S.3) and (S.7), we obtain

$$\Delta_{3\mathrm{D}}^2 = \Delta_0^2 - 4\mathcal{A}DJ_a. \quad (\mathrm{S}.10)$$

Using the critical coupling $J_a^{\text{crit}}$ as defined for given $D$ and $J_c$ by (S.8), we can rewrite this result in an alternative form:

$$\Delta_{3D}^2 = \Delta_0^2 - \Delta_0^2 \frac{J_a}{J_a^{\text{crit}}}. \qquad (S.11)$$

This is the Eq. (6) from the main text.

## V. FITS TO THE EXPERIMENTAL PHASE BOUNDARIES

A given set of DTN spin-Hamiltonian parameters can be represented by a point in the Sakai–Takahashi phase diagram plane (Fig. 7(d) of the main text, where we denote such points by the purple open circles) [11]. The Hamiltonian parameters $D$ and $J_c$ are changing with pressure, so the corresponding point on the diagram also changes its position. Since the change in $D$ and $J_c$ is linear, and $J_a$ is a constant, the resulting trajectory in this "phase space" is nearly a straight line. Such line may have only a single interception with the curved phase boundary between the gapped and long-range ordered states. The slope of the line that represents the trajectory is given by the ratio between $\partial D/\partial P$ and $\partial J_c/\partial P$. Since the experimentally known pressure dependence of $H_{c2}$ locks the linear relationship between these quantities, finding this slope is effectively a single-parameter problem. The critical pressure $P_c$ is then ideally suited to play the role of this open parameter. The parameterized trajectory we are discussing is crossing the known phase boundary at $P = P_c$, which allows the evaluation of the corresponding $\partial D/\partial P$ and $\partial J_c/\partial P$ values. Pairs of parameters, obtained for different choices of $P_c$, are shown in Fig. S5b in the magnetostriction notation (see the next section). Since the value of $P_c$ remains *the only parameter* in the description of the experimental data, we can utilize a numeric criterion to optimize the agreement. As such criterion, we consider the mean-square deviation

$$\chi^2 = \mu_0 \sqrt{\sum \left(H_{c1}^{\text{obs}} - H_{c1}^{\text{calc}}\right)^2}, \qquad (S.12)$$

with the calculated $H_{c1}$ values obtained from our *ansatz*, and $H_{c1}^{\text{obs}}$ being the experimental data. This quantity is shown in Fig S5a. The deviation is robustly minimized in the range $P_c = 4.2 \pm 0.3$ kbar, in very good agreement with the "apparent" value of the critical pressure.

## VI. ELASTIC CONSTANTS

The pressure dependencies of $D$ and $J_c$ we have found can be expressed as magnetostriction coefficients. However, under a hydrostatic-pressure conditions the relation between the length reduction and the Young's modulus for the given direction is not straightforward. Here we derive the corresponding equations prior to discussing the measured lattice-parameters changes under pressure.

### A. Hydrostatic pressure vs uniaxial strain

We need to consider the elasticity-theory equations for three different cases: hydrostatic pressure, uniaxial strain along the $x$ direction, and uniaxial strain along $z$. Since we are dealing with a tetragonal crystal, we can neglect the shear components. Than, the equations that define the relation between the stress $\sigma$ and the strain $\varepsilon$ are reduced to a 3-by-3 matrix equation:

$$\begin{pmatrix} \sigma_x \\ \sigma_y \\ \sigma_z \end{pmatrix} = \mathbb{C} \begin{pmatrix} \varepsilon_x \\ \varepsilon_y \\ \varepsilon_z \end{pmatrix}. \qquad (S.13)$$

Here, the elasticity matrix $\mathbb{C}$ is composed of several symmetry-allowed elastic moduli:

$$\mathbb{C} = \begin{pmatrix} c_{xx} & c_{xy} & c_{xz} \\ c_{xy} & c_{xx} & c_{xz} \\ c_{xz} & c_{xz} & c_{zz} \end{pmatrix}. \qquad (S.14)$$

The inverse problem $\vec{\varepsilon} = \mathbb{S}\vec{\sigma}$ is formulated with the help of the compliance matrix $\mathbb{S} = \mathbb{C}^{-1}$, which has the same symmetry. The compliance-matrix components are [12]:

$$\begin{aligned} s_{xx} &= \frac{c_{xx}c_{zz} - c_{xz}^2}{c_0(c_{xx} - c_{xy})}, \\ s_{zz} &= \frac{c_{xx} + c_{xy}}{c_0}, \\ s_{xy} &= \frac{c_{xz}^2 - c_{xy}c_{zz}}{c_0(c_{xx} - c_{xy})}, \\ s_{xz} &= -\frac{c_{xz}}{c_0}, \end{aligned} \qquad (S.15)$$

with $c_0 = c_{zz}(c_{xx} + c_{xy}) - 2c_{xz}^2$.

For a uniaxial stress applied along the $x$ direction, $\vec{\sigma} = (P, 0, 0)$. Thus, the strain along that direction is:

$$\varepsilon_x = s_{xx}P.$$

For the $z$ direction, with $\vec{\sigma} = (0, 0, P)$ we correspondingly have

$$\varepsilon_z = s_{zz}P.$$

This gives us the conventional Young's moduli

$$E_{xx} = s_{xx}^{-1} = c_{xx} - c_{xy} + \frac{(c_{zz}c_{xy} + c_{xz}^2)(c_{xx} - c_{xy})}{c_{xx}c_{zz} - c_{xz}^2}, \qquad (S.16)$$

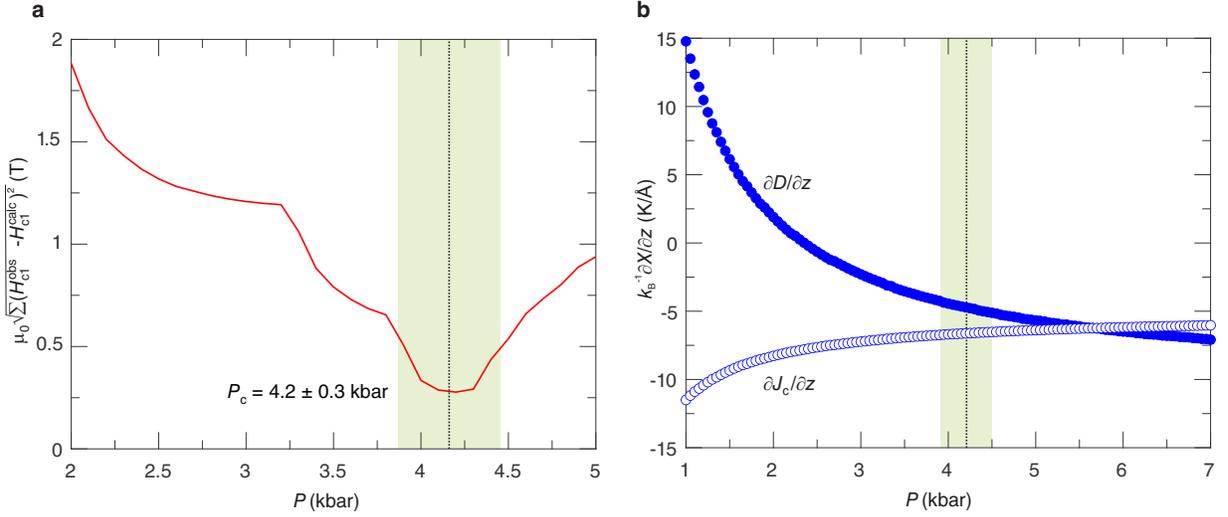

FIG. S5. The model results depending on the selected critical pressure value $P = P_c$. **a** Mean-square deviation between the RPA *ansatz* fit and the actually observed values of $H_{c1}$ as a function of pressure that we assign as critical. The agreement is optimized around $P_c \simeq 4.2$ kbar; the error bar represents the approximate threshold at which the $\chi^2$ (see Eq. (S.12)) doubles compared to its minimal value. **b** The magnetostriction coefficients as a function of pressure that we assign as critical. The optimal region is highlighted.

and

$$E_{zz} = s_{zz}^{-1} = c_{zz} - 2\frac{c_{xz}^2}{c_{xx} + c_{xy}}. \quad (S.17)$$

Under hydrostatic pressure the compression is different. The stress is $\vec{\sigma} = (P, P, P)$, and the strain is $\vec{\varepsilon} = (\varepsilon_x, \varepsilon_x, \varepsilon_z)$. For the strain components, we find:

$$\varepsilon_x = (s_{xx} + s_{xy} + s_{xz})P = \frac{c_{zz} - c_{xz}}{c_{zz}(c_{xx} + c_{xy}) - 2c_{xz}^2}P,$$
$$\varepsilon_z = (2s_{xz} + s_{zz})P = \frac{c_{xx} + c_{xy} - 2c_{xz}}{c_{zz}(c_{xx} + c_{xy}) - 2c_{xz}^2}P. \quad (S.18)$$

Hence, under the hydrostatic conditions our pressure-to-strain ratios are:

$$\left[\frac{\Delta P}{\Delta a/a}\right] = \frac{c_{zz}(c_{xx} + c_{xy}) - 2c_{xz}^2}{c_{zz} - c_{xz}}, \quad (S.19)$$

$$\left[\frac{\Delta P}{\Delta c/c}\right] = \frac{c_{zz}(c_{xx} + c_{xy}) - 2c_{xz}^2}{c_{xx} + c_{xy} - 2c_{xz}}. \quad (S.20)$$

Note, that from Eqs. (S.17) and (S.20) follows $\left[\frac{\Delta P}{\Delta c/c}\right] = E_{zz}\frac{c_{xx} + c_{xy}}{c_{xx} + c_{xy} - 2c_{xz}}$. We obtain a significant boost in the compression compared to the uniaxial case due to off-diagonal components of the elasticity matrix.

### B. Experimental compressibility

As demonstrated above, the hydrostatic compressibility modulus for the $c$ direction is:

$$\left[\frac{\Delta P}{\Delta c/c}\right] = E_{zz}\frac{c_{xx} + c_{xy}}{c_{xx} + c_{xy} - 2c_{xz}}. \quad (S.21)$$

Here, $E_{zz}$ is the true Young's modulus for longitudinal stress, and $c_{\alpha\beta}$ are the various elastic moduli. According to Ref. [13], these moduli are $c_{xx} = 26.1$ GPa, $c_{xy} = 15.3$ GPa, and $c_{xz} = 12.4$ GPa, yielding a factor 4.76 in Eq. (S.21). We find $\left[\frac{\Delta P}{\Delta c/c}\right] = 23.1 \pm 1.7$ GPa, hence $E_{zz} = 4.9 \pm 0.4$ GPa. This is reasonably close to the value $7.5 \pm 0.7$ GPa [13]. Possible systematic errors in the determination of the elastic moduli $c_{\alpha\beta}$ would affect either estimate.

### C. Magnetostriction coefficients

Using the measured lattice parameter dependence and derived $\partial D/\partial P$, $\partial J_c/\partial P$, we can estimate the magnetostriction coefficients. Let $z$ parameterize the bond length along the $c$ direction, then:

$$\frac{\partial X}{\partial z} = \frac{\partial X}{\partial P}\left(\frac{\partial c}{\partial P}\right)^{-1}.$$

Within this model, the magnetostriction coefficients $k_B^{-1}\partial J_c/\partial z$ and $k_B^{-1}\partial D/\partial z$ as function of critical pressure choice are shown in Fig. S5b. The possible values within the optimal $P_c$ range are highlighted; $k_B^{-1}\partial D/\partial z = -4.1 \pm 0.7$ K/Å and $k_B^{-1}\partial J_c/\partial z = -6.5 \pm 0.3$ K/Å. Surprisingly, they noticeably deviate from the set of magnetostriction coefficients (0 and $-2.5$ K/Å) previously

found at the zero pressure [13]. The latter values are clearly insufficient to describe the rapid growth of $H_{c2}$ under pressure, yielding only 0.25 T/kbar (which is about 1/3 of the actual slope).

## VII. HIGH-PRESSURE ESR

### A. GSWT description

The key idea of generalized spin wave theory (GSWT) [6, 14] is to approximate the collective ground state of a magnetic system as a simple product of single-ion states. Both, quantum-disordered states (products of local singlets $|0\rangle_i$) and magnetic states with $\left\langle \hat{\mathbf{S}}_i \right\rangle \neq 0$ (products of general combinations of $|0\rangle_i, |-1\rangle_i, |+1\rangle_i$) are examples of such states. The particular mixing weights in such combinations can be found from minimizing the mean-field energy. This is one of the crucial differences to the linear spin wave theory (LSWT) where a ground state with $|\langle \hat{\mathbf{S}}_i \rangle| = S$ is postulated. Within GSWT, arbitrary values of $|\langle \hat{\mathbf{S}}_i \rangle| \leq S$ are possible, partially accounting for the effect of quantum fluctuations.

Having identified the mean-field ground state on every ion, the remaining two states of the appropriate single-spin basis sets remain as excited states. It is now possible to rewrite the Hamiltonian in the second-quantization notation associating the bosonic particle creation operators $\hat{a}_i^\dagger, \hat{b}_i^\dagger$ with creating these local excited states. This is another difference to LSWT that technically accounts only for the first excited state. By Fourier transformation of the bosonic operators we can obtain the second-quantization Hamiltonian in the momentum space representation. Then it can be diagonalized by Bogolyubov transformation, yielding the dispersion relations for the possible excitation branches. We handle these calculations numerically for the given Hamiltonian parameters using our own code. Then the excitations at $q = 0$ are the ones of interest for ESR experiments.

We would also like to note that the recently released Sunny library (for simulating SU($N$) dynamics in systems of complex magnetic ions) [15] utilizes the same approach and produces identical results for the case of DTN.

### B. ESR frequency-field diagrams

The method outlined above allows us to calculate the frequency-field dependence of the uniform, $q = 0$ excitations for DTN at different pressures. Three modes (labeled as A, B, and C) are the "collective" descendants of $|0\rangle \to |\mp 1\rangle$ single-ion transitions at low fields, or the $|+1\rangle \to |0\rangle$ transition at high fields.

In Fig. S6, we show the frequency-field diagrams for the uniform magnetic excitations obtained from ESR experiments at all measured pressures. We also include the reference zero-pressure data from Ref. [16]. The solid lines in Fig. S6 represent the predicted zero-temperature ESR single-magnon transitions for the DTN spin Hamiltonian at corresponding pressures.

Qualitatively the weakly changing spectra are consistent with the above analysis (e.g., pressure-induced shift to the right for the branch B, and to the left for the branch A in both theory and experiment). However, the description of the spectrum in the field range, corresponding to the low-$T$, long-range ordered-phase is not accurate. At zero pressure, the central part of the spectrum is known to show a strong temperature dependence between 2 and 0.5 K [1], eventually matching the theoretical prediction at $T = 0$. The mismatch appears to be result of the simultaneous action of thermal and quantum fluctuations, not fully accounted for by GSWT. The case of DTN under pressure would be an excellent benchmark for the possible future extensions of the theory incorporating such corrections.

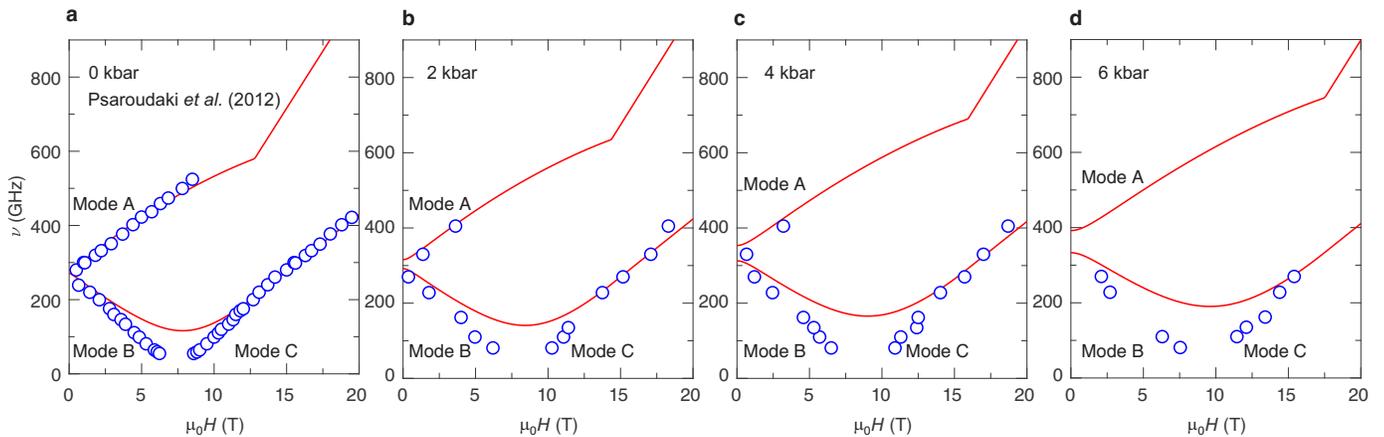

FIG. S6. Frequency–field diagram of the spin excitations in DTN under different pressures. **a** Zero-pressure $T \simeq 2$ K data taken from Ref. [16] (open circles). Solid lines correspond to a $T = 0$ GSWT description of the spin dynamics. **b** Data at $T \simeq 4$ K and 2 kbar pressure (open circles), 2 kbar GSWT calculations (red lines). **c** Same for 4 kbar. **d** Same for 6 kbar.



\end{document}